\documentclass[sigconf,dvipsnames,natbib=true]{acmart}

\usepackage[utf8]{inputenc}
\usepackage{booktabs,multirow}

\usepackage{dsfont}
\usepackage{bbm}
\usepackage{arydshln}
\usepackage{hyperref}

\hypersetup{
    colorlinks=true,
    linkcolor=blue,
    filecolor=magenta,      
    urlcolor=cyan,
    }

\usepackage{color}
\usepackage{rotating}
\usepackage{amsmath}
\usepackage{array}
\usepackage{enumitem}
\usepackage[linesnumbered,algoruled,boxed,lined]{algorithm2e}
\usepackage{kotex}
\usepackage{balance}
\usepackage[short]{optidef}
\usepackage{graphicx}
\usepackage{adjustbox}

\usepackage{contour}
\usepackage[normalem]{ulem}

\contourlength{0.8pt}

\newcolumntype{P}[1]{>{\centering\arraybackslash}p{#1}}
\newcolumntype{R}[1]{>{\RaggedLeft\arraybackslash}p{#1}}

\newcommand{\ie}{{\it i.e.}}
\newcommand{\eg}{{\it e.g.}}

\newcommand{\ul}{\underline}{}

\newtheorem{theorem}{Theorem}[section]
\newtheorem{lemma}{Lemma}[section]

\newcommand{\red}[1]{\textbf{\textcolor{red}{#1}}}
\newcommand{\blue}[1]{\ul{\textcolor{blue}{#1}}}

\copyrightyear{2025}
\acmYear{2025}
\setcopyright{cc}
\setcctype{by-nc-nd}
\acmConference[SIGIR '25]{Proceedings of the 48th International ACM SIGIR Conference on Research and Development in Information Retrieval}{July 13--18, 2025}{Padua, Italy}
\acmBooktitle{Proceedings of the 48th International ACM SIGIR Conference on Research and Development in Information Retrieval (SIGIR '25), July 13--18, 2025, Padua, Italy}
\acmISBN{979-8-4007-1592-1/2025/07}
\acmDOI{10.1145/3726302.3730116}

\begin{document}

\title{Why is Normalization Necessary for Linear Recommenders?}

\author{Seongmin Park}
\affiliation{
  \institution{Sungkyunkwan University}
  \city{Suwon}
  \country{Republic of Korea}}
\email{psm1206@skku.edu}

\author{Mincheol Yoon}
\affiliation{
  \institution{Sungkyunkwan University}
  \city{Suwon}
  \country{Republic of Korea}}
\email{yoon56@skku.edu}

\author{Hye-young Kim}
\affiliation{
  \institution{Sungkyunkwan University}
  \city{Suwon}
  \country{Republic of Korea}}
\email{khyaa3966@skku.edu}

\author{Jongwuk Lee}\authornote{Corresponding author}
\affiliation{
  \institution{Sungkyunkwan University}
  \city{Suwon}
  \country{Republic of Korea}}
\email{jongwuklee@skku.edu}

\ccsdesc[500]{Information systems~Recommender systems}

\keywords{Collaborative filtering; linear autoencoders; normalization; popularity bias; neighborhood bias}

\begin{abstract}

Despite their simplicity, linear autoencoder (LAE)-based models have shown comparable or even better performance with faster inference speed than neural recommender models. However, LAEs face two critical challenges: (i) \emph{popularity bias}, which tends to recommend popular items, and (ii) \emph{neighborhood bias}, which overly focuses on capturing local item correlations. To address these issues, this paper first analyzes the effect of two existing normalization methods for LAEs, \ie, \emph{random-walk} and \emph{symmetric normalization.} Our theoretical analysis reveals that normalization highly affects the degree of popularity and neighborhood biases among items. Inspired by this analysis, we propose a versatile normalization solution, called \emph{\textbf{D}ata-\textbf{A}daptive \textbf{N}ormalization (\textbf{DAN})}, which flexibly controls the popularity and neighborhood biases by adjusting item- and user-side normalization to align with unique dataset characteristics. Owing to its model-agnostic property, DAN can be easily applied to various LAE-based models. Experimental results show that DAN-equipped LAEs consistently improve existing LAE-based models across six benchmark datasets, with significant gains of up to 128.57\% and 12.36\% for long-tail items and unbiased evaluations, respectively.
Refer to our code in \url{https://github.com/psm1206/DAN}.

\end{abstract}


\maketitle

\section{Introduction}\label{sec:introduction}

Collaborative filtering (CF)~\cite{GoldbergNOT92CF, HerlockerKBR99CFUserKNN} is the dominant solution for developing recommender systems because it uncovers hidden collaborative signals from user-item interactions. Existing CF models can be categorized into \emph{linear} and \emph{non-linear} approaches depending on how user/item correlations are learned. With the success of deep learning, a lot of non-linear CF models have employed various neural architectures, including autoencoders~\cite{WuDZE16CDAE, LiangKHJ18MultVAE, ShenbinATMN20RecVAE, LobelLGC20RaCT, YeX023} (AEs), recurrent neural networks (RNNs)~\cite{HidasiKBT15GRU4Rec, LiRCRLM17NARM}, graph neural networks~\cite{0001DWLZ020LightGCN, Wang0WFC19NGCF, 0002JP21LTOCF, ShenWZSZLL21GFCF, MaoZXLWH21, ChoiHPC23BSPM} (GNNs), and transformers~\cite{KangM18SASRec, SunLWPLOJ19BERT4Rec, Shin0WP24}.

Although these non-linear models initially showed promising results through complex user/item relationships, recent studies~\cite{DacremaCJ19, SunY00Q0G20, abs-1905-01395, DacremaBCJ21, abs-2305-01801, ChoiHPC23BSPM, PengLSM24sgfcf} have revealed that the linear models can achieve competitive or even significant gains over the non-linear models. This is because linear models are less prone to overfitting in learning sparse user-item interactions. Also, their computational efficiency enables rapid adoption in real-world applications. In this sense, this paper focuses on the linear models using item neighborhoods, also known as \emph{linear autoencoders (LAEs)}~\cite{NingK11SLIM, Steck19EASE, Steck20edlae, SteckDRJ20ADMMSLIM, JeunenBG20CEASE, SteckL21Higher, VancuraAKK22ELSA}.

Formally, LAEs operate on a user-item interaction matrix $\mathbf{X} \in \{0,1\}^{m \times n}$ for $m$ users and $n$ items, learning an \emph{item-to-item weight matrix} $\mathbf{B} \in \mathbb{R}^{n \times n}$ to reconstruct the original matrix $\mathbf{X}$ from matrix multiplication $\mathbf{X} \cdot \mathbf{B}$. Conceptually, $\mathbf{B}$ denotes a single hidden layer to act as both an encoder and a decoder for the input matrix $\mathbf{X}$. Existing studies~\cite{NingK11SLIM, Steck19EASE, Steck20edlae, SteckDRJ20ADMMSLIM, JeunenBG20CEASE, SteckL21Higher, VancuraAKK22ELSA} formulate LAEs as a convex optimization with additional constraints, yielding a closed-form solution while avoiding a trivial solution, \ie, $\hat{\mathbf{B}} = \mathbf{I}$. Notably, the representative LAE-based models, such as EASE$^\text{R}$~\cite{Steck19EASE}, DLAE/EDLAE~\cite{Steck20edlae}, and RLAE/RDLAE~\cite{MoonKL23RDLAE}, have shown state-of-the-art performance results on large-scale datasets, \eg, ML-20M, Netflix, and MSD.

However, existing LAEs face two challenges.
\textbf{(C1) Popularity bias}: Inevitably, the learned weight matrix $\hat{\mathbf{B}}$ is heavily influenced by popular items, leading to popular items being excessively recommended to users~\cite{0007D0F0023, Zhu0ZZWC21, GuptaGMVS2019, SaitoYNSN20, WeiFCWYH21}. \textbf{(C2) Neighborhood bias}: It refers to the tendency to overly focus on local item relationships that capture individual user preferences, predominantly from a few highly engaged users~\cite{MuLZZY24}.
Since LAEs directly operate on the input matrix $\mathbf{X}$, they primarily capture these local relationships while failing to discover global item relationships shared by most users, which are crucial for capturing principal CF patterns.

To address these challenges, we delve into normalizing the user-item matrix $\mathbf{X}$ for LAEs. As the conventional technique, normalization has been widely adopted in numerous linear and non-linear recommendation models~\cite{Wang0WFC19NGCF, 0001DWLZ020LightGCN, YuXCCHY24XSimGCL, ChoiHPC23BSPM, ShenWZSZLL21GFCF, ParkSS2024turbocf, HongCLKP24svdae}.
Despite its widespread usage, there has been a lack of in-depth analysis exploring the underlying effects of normalization, particularly in the context of linear recommenders. This motivates us to ask the following key questions about the normalization for LAEs: (i) \emph{How do we apply normalization to LAEs?} (ii) \emph{How does normalization affect popularity bias?} (iii) \emph{How does normalization influence neighborhood bias?}

Firstly, we examine how to apply normalization to LAEs by exploring item- and user-side normalizations on the reconstruction term $\|\mathbf{X} - \mathbf{X} \mathbf{B}\|_F^2$ and the regularization term $\|\mathbf{B}\|_F^2$. For item normalization, we employ a diagonal matrix $\mathbf{D}_{I} \in \mathbb{R}^{n \times n}$ to control the row- and column-wise importance of items in forming the weight matrix $\mathbf{B}$. For user normalization, we adopt a diagonal matrix $\mathbf{D}_U \in \mathbb{R}^{m \times m}$ to adjust user-wise importance in reconstructing $\mathbf{X}$. These normalizations directly affect the gram matrix $\mathbf{P} = \mathbf{X}^{\top}\mathbf{X} \in \mathbb{R}^{n \times n}$, which determines item co-occurrences in LAE's closed-form solution $\hat{\mathbf{B}} = \left(\mathbf{P} + \lambda \mathbf{I}\right)^{-1}\mathbf{P}$. We analyze these effects through existing methods such as \emph{random-walk} and \emph{symmetric normalization}. 
While these approaches handle item/user importance, they both suffer from a fundamental limitation of using fixed weights.

\begin{figure}
\begin{tabular}{cc}
\includegraphics[height=3.0cm]{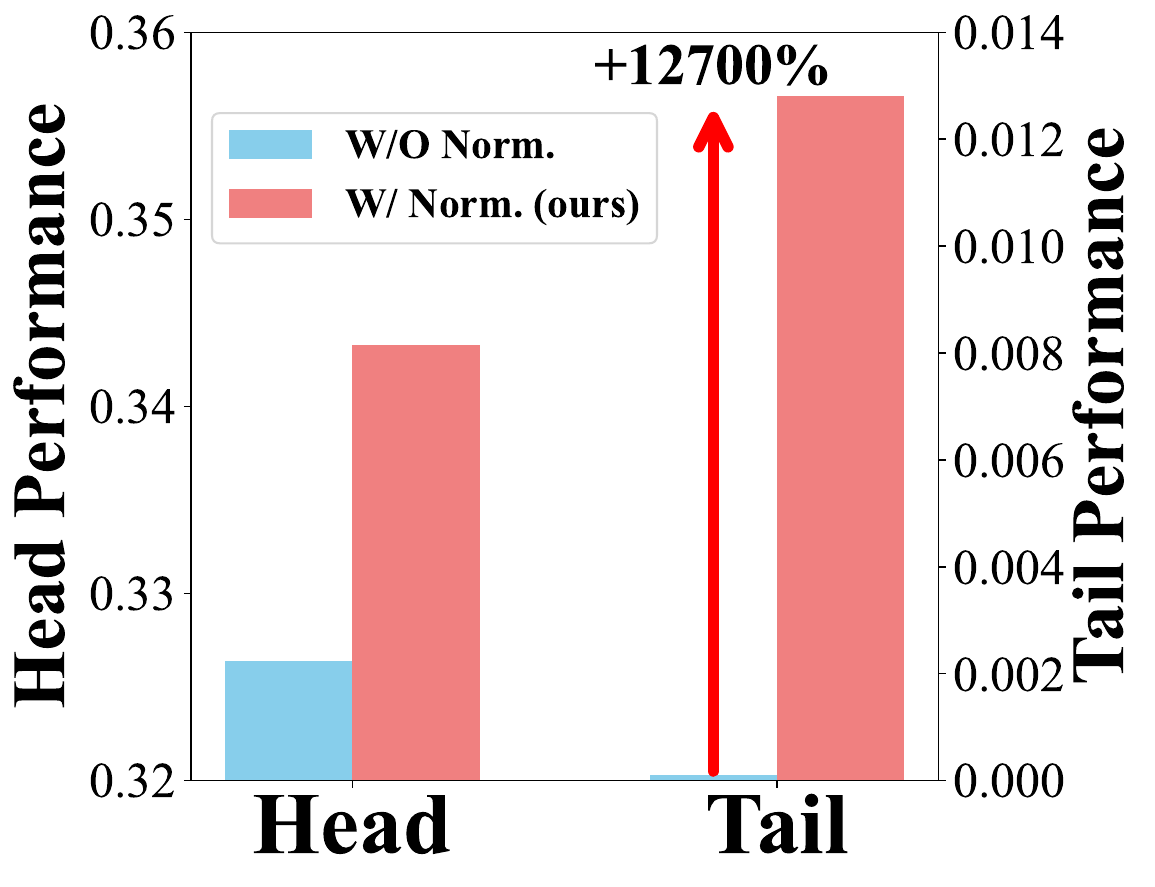} &
\includegraphics[height=3.0cm]{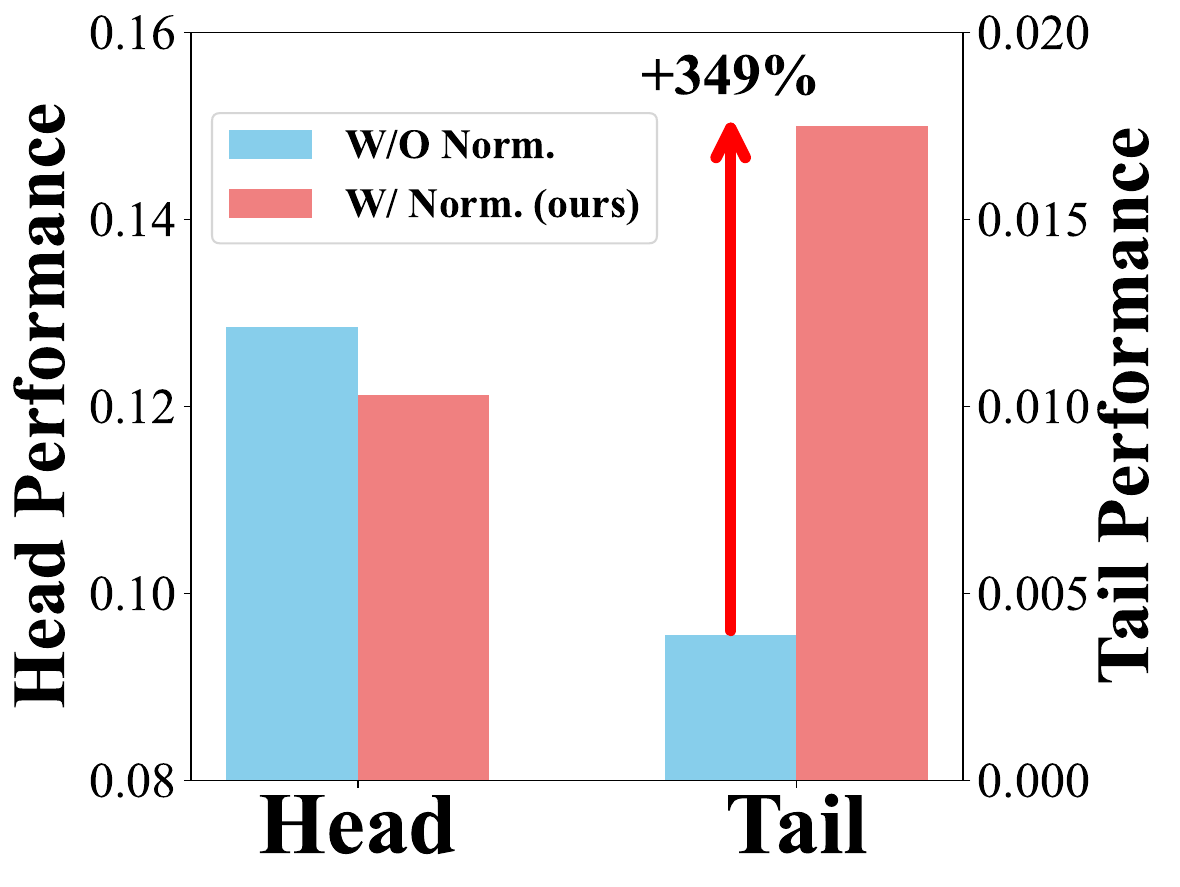} \\
\multicolumn{1}{c}{(a) ML-20M} &\multicolumn{1}{c}{(b) Yelp2018}
\vspace{-2mm}
\end{tabular}
\caption{Performance of popular and unpopular items on ML-20M and Yelp2018. The x-axis categorizes items into `Head' (top 20\% popular items) and `Tail' (the remaining items), while the y-axis represents NDCG@20. `W/O Norm.' and `W/ Norm.' denote LAE without and with normalization.}\label{fig:head_tail_perform_norm}
\vspace{-3mm}
\end{figure}

Secondly, we conduct an empirical study to analyze the impact of normalization on popularity bias. Figure~\ref{fig:head_tail_perform_norm} illustrates the performance for both popular (head) and unpopular (tail) items. LAEs without normalization (\ie, W/O Norm.) predominantly recommend popular items, leading to high performance for head items but low performance for tail items. In contrast, LAEs with normalization (\ie, W/ Norm.) substantially improve Tail performance across datasets while maintaining competitive Head performance. It reveals that normalization effectively adjusts the degree of popularity bias.

\begin{figure}
\begin{tabular}{cc}
\includegraphics[height=3.0cm]{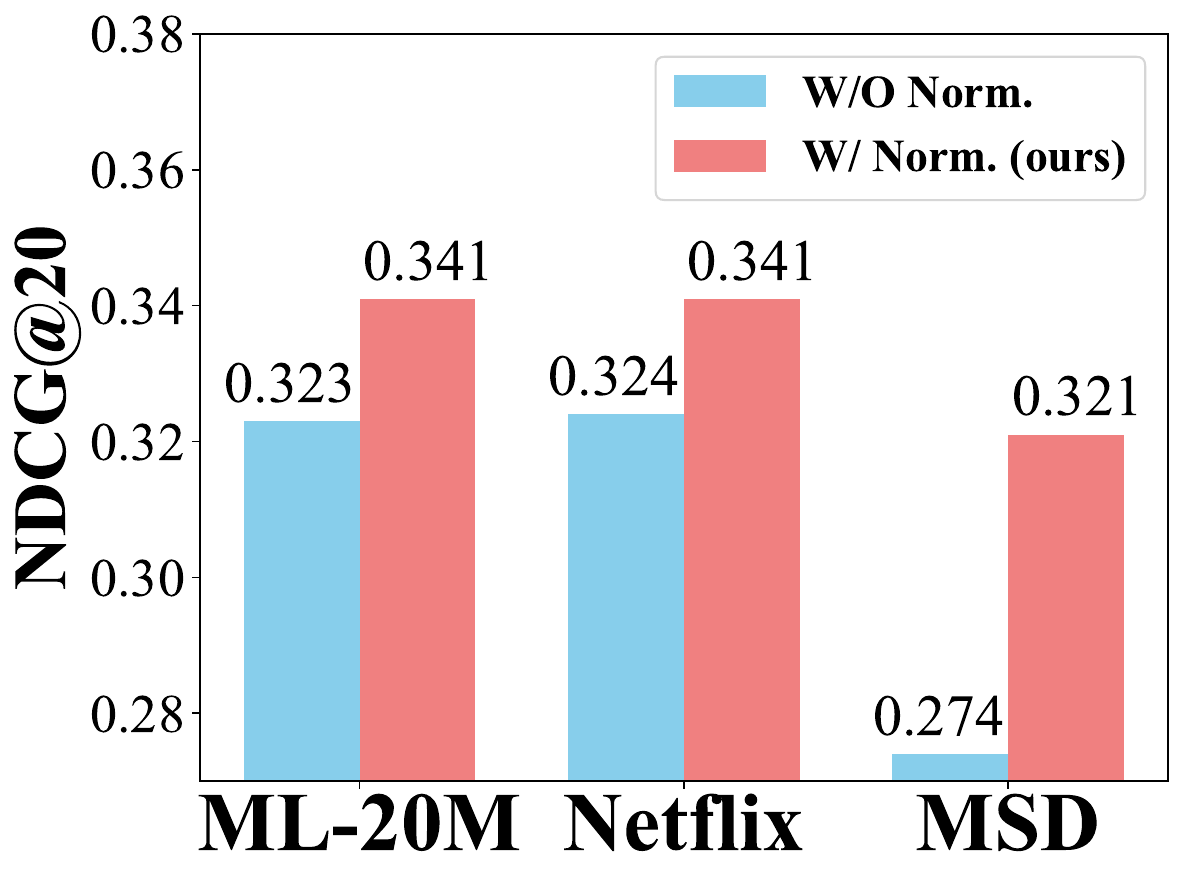} &
\includegraphics[height=3.0cm]{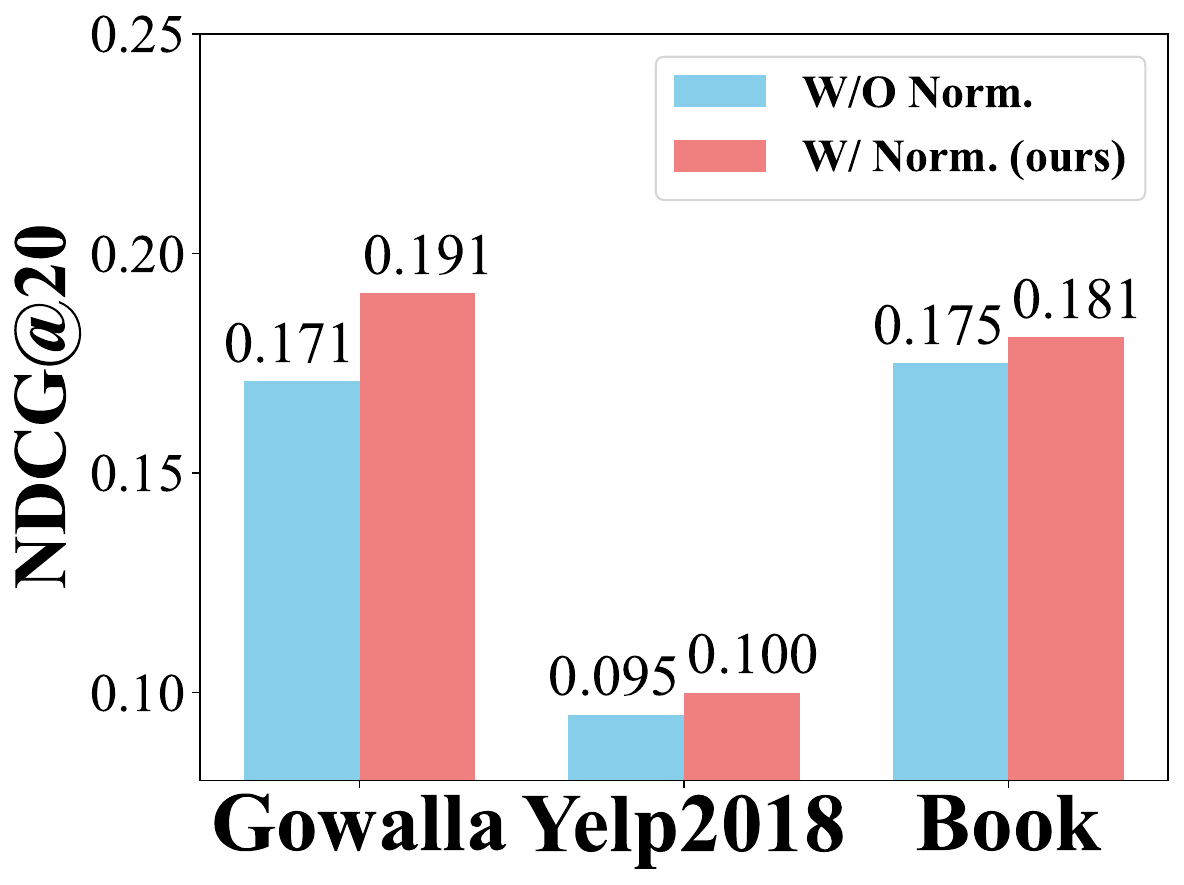} \\
\multicolumn{1}{c}{(a) High-homophilic group} &\multicolumn{1}{c}{(b) Low-homophilic group}
\vspace{-2mm}
\end{tabular}
\caption{Performance for six datasets categorized into two groups: High-homophilic group (ML-20M, Netflix, and MSD) with low neighborhood bias and Low-homophilic group (Gowalla, Yelp2018, and Amazon-book) with high neighborhood bias. The x-axis lists the individual datasets. }\label{fig:dataset_group_perform_norm}
\vspace{-3mm}
\end{figure}

We lastly investigate the normalization effect by categorizing six datasets into high- and low-homophilic groups. High-homophilic datasets have densely connected similar items, indicating shared global patterns. (Detailed homophily metric is discussed in Section~\ref{sec:model}.) As depicted in Figure~\ref{fig:dataset_group_perform_norm}, normalization consistently improves LAEs across datasets by capturing global item correlations while mitigating local relationships. This is especially evident in high-homophilic datasets, which show substantial performance gains. They also show high absolute performance due to their inherent item relationships that align with CF principles. These findings demonstrate that normalization is a key mechanism for balancing global and local item relationships, modulating neighborhood bias.

Based on these findings, we propose a simple yet effective normalization, called \emph{\textbf{D}ata-\textbf{A}daptive \textbf{N}ormalization (\textbf{DAN})}, effectively balancing the popularity of items/users depending on dataset-specific characteristics. It comprises two key components: (i) \textbf{Item-adaptive normalization} modulates the strength of item popularity by the skewness of item distributions, where we theoretically demonstrate its effectiveness in adjusting popularity bias. (ii) \textbf{User-adaptive normalization} adjusts the influence of users by the newly proposed homophily metric tailored for recommendations, which helps control neighborhood bias by discovering meaningful global relationships among items. Thanks to its model-agnostic property, DAN can be easily adapted to various LAE-based models. Experimental results show that DAN-equipped LAEs outperform existing normalization methods and state-of-the-art models on six benchmark datasets. Notably, it achieves up to 128.57\% and 12.36\% performance gains for long-tail items and unbiased evaluation, respectively.

The key contributions of this paper are summarized as follows.
\begin{itemize}[leftmargin=5mm]
    \item \textbf{Mathematical analysis}: We are the first to investigate existing normalization methods on LAEs. It is observed that existing normalization methods can alleviate the popularity bias and neighborhood bias, but they are limited in handling the unique characteristics of the datasets. (Section~\ref{sec:analysis})

\vspace{0.5mm}
    \item \textbf{Data-customized normalization}: We propose simple yet effective normalization \textbf{DAN}, which adjusts the degree of popularity and neighborhood biases by the dataset characteristics, such as the skewness of item distributions (\eg, item-side Gini index) and the weighted homophily ratio. (Section~\ref{sec:model})
    
\vspace{0.5mm}
    \item \textbf{Extensive validation}: We demonstrate that DAN-equipped LAEs achieve superior performance over fourteen existing CF models on six datasets. We also conduct experiments to evaluate DAN's efficiency, showing that it adds negligible computational cost compared to existing LAEs. (Sections~\ref{sec:setup}--\ref{sec:results})

\end{itemize}

\section{Preliminaries}\label{sec:background}

\subsection{Linear Autoencoders (LAEs)}

Assuming implicit user feedback, we represent the user-item interaction matrix $\mathbf{X}$ as a binary matrix, \ie, $\mathbf{X} \in \{0,1\}^{m \times n}$. If a user $u$ has interacted with an item $i$, then $\mathbf{X}_{ui}=1$; otherwise, $\mathbf{X}_{ui}=0$.

In this paper, we mainly focus on addressing LAE-based recommender models~\cite{NingK11SLIM, Steck19EASE, Steck20edlae, JeunenBG20CEASE, SteckL21Higher, VancuraAKK22ELSA}, to learn an \emph{item-to-item weight matrix} $\mathbf{B} \in \mathbb{R}^{n \times n}$. Specifically, the objective function of LAEs is formulated by the reconstruction error and L2 regularization.
\begin{equation}\label{eq:lae_objective}
  \min_{\mathbf{B}} \|\mathbf{X} - \mathbf{X} \mathbf{B}\|_F^2 + \lambda \|\mathbf{B}\|_F^2,
\end{equation}
\noindent where $\lambda$ is the hyperparameter to control the regularization in $\mathbf{B}$. When $\lambda=0$, it becomes a trivial solution, \ie, $\hat{\mathbf{B}}_{LAE} = \mathbf{I}$.

The closed-form solution of LAE is easily derived as follows.
\begin{equation}\label{eq:lae_solution}
\hat{\mathbf{B}}_{LAE} = \left(\mathbf{X}^\top \mathbf{X} + \lambda \mathbf{I}\right)^{-1}\left(\mathbf{X}^\top \mathbf{X}\right) = \left(\mathbf{P} + \lambda \mathbf{I}\right)^{-1} \mathbf{P}.
\end{equation}

Here, $\mathbf{P} = \mathbf{X}^{\top}\mathbf{X} \in \mathbb{R}^{n \times n}$ represents a gram matrix for items. It is symmetric, and the entry $\mathbf{P}_{ij}$ indicates the co-occurrence between two items $i$ and $j$ by all users.
\begin{equation}\label{eq:grammatrix}
\mathbf{P}_{ij}= \left(\mathbf{X}^{\top}\mathbf{X}\right)_{ij} = freq(i,j).
\end{equation}

The gram matrix $\mathbf{P}$ tends to be biased in two aspects: (i) it reflects only raw co-occurrence between items, leading to the \emph{popularity bias} in $\hat{\mathbf{B}}_{LAE}$, and (ii) it considers only direct item relationships, resulting in the \emph{neighborhood bias}, indicating that the global item correlations are non-trivial to reflect.

For inference, the prediction score $s_{ui}$ for the item $i$ to the user $u$ is computed as follows.
\begin{equation}\label{eq:prediction}
    s_{ui} = \mathbf{X}_{u*} \cdot \hat{\mathbf{B}}_{*i},
\end{equation}
where $\mathbf{X}_{u*}$ and $\hat{\mathbf{B}}_{*i}$ are the row vector for the user $u$ in $\mathbf{X}$ and the column vector for the item $i$ in $\mathbf{B}$, respectively. That is, each column in $\hat{\mathbf{B}}$ means a \emph{target item} to be recommended to the user. Since item $i$ in the user history is multiplied by the $i$-th row in $\hat{\mathbf{B}}$, each row in $\hat{\mathbf{B}}$ represents a \emph{source item} the user interacts with. We can also interpret the rows and columns of the gram matrix $\mathbf{P}$ as the source and target items because $\hat{\mathbf{B}}$ is composed of $\mathbf{P}$ in Eq.~\eqref{eq:lae_solution}.

\subsection{Existing Normalization Methods}
This section introduces two representative normalization methods, such as \emph{random-walk (RW)} and \emph{symmetric (Sym) normalization}, commonly used in recommendation models~\cite{CooperLRS14P3alpha, ChristoffelPNB15, ShenWZSZLL21GFCF, Wang0WFC19NGCF, 0001DWLZ020LightGCN, ChoiHPC23BSPM}.

\vspace{1mm}
\noindent
\textbf{RW normalization}. Let $\mathcal{U}$ and $\mathcal{I}$ be a set of $m$ users and $n$ items, respectively. User-item interactions are interpreted by a user-item bipartite graph $\mathcal{G}$ = ($\mathcal{U} \cup \mathcal{I}$, $\mathcal{E}$), where nodes represent users and items. Here, an edge $e \in \mathcal{E}$ indicates the interaction between the user and the item. Let $\mathbf{A}$ denote an adjacency matrix representing the user-item interactions on the graph $\mathcal{G}$. By considering the random walk movement, the transition probability is computed by normalizing the number of either users or items.
\begin{equation} \label{eq:transition}
    \tilde{\mathbf{A}} = \mathbf{D}^{-1} \mathbf{A} = 
    \begin{bmatrix}
    \mathbf{D}_U^{-1} & 0 \\
    0 & \mathbf{D}_I^{-1} \\
    \end{bmatrix}
    \begin{bmatrix}
    0 & \mathbf{X} \\
    \mathbf{X}^{\top} & 0 \\
    \end{bmatrix} = 
    \begin{bmatrix}
    0 & \mathbf{D}_U^{-1}\mathbf{X} \\
    \mathbf{D}_I^{-1}\mathbf{X}^{\top} & 0 \\
    \end{bmatrix},
\end{equation}
where $\mathbf{D} \in \mathbb{R}^{(m+n) \times (m+n)}$ is a diagonal degree matrix. $\mathbf{D}_U \in \mathbb{R}^{m \times m}$ represents a diagonal degree matrix for users, where each entry $\mathbf{D}_{uu}$ is the number of items interacted by the user $u$, \ie, $\mathbf{D}_{uu} = \sum_{i=1}^{n}{\textbf{X}_{ui}}$, and vice versa for items with $\mathbf{D}_I \in \mathbb{R}^{n \times n}$ where $\mathbf{D}_{ii} = \sum_{u=1}^{m}{\textbf{X}_{ui}}$.

Two distinct matrices $\mathbf{D}_U^{-1}\mathbf{X}$ and $\mathbf{D}_I^{-1}\mathbf{X}^{\top}$ are the user-to-item and item-to-user transition probabilities, respectively.
They are interpreted as user/item normalization.
\begin{equation} \label{eq:x_rw_norm}
    \tilde{\mathbf{X}}_{rw}^{(user)} = \mathbf{D}_U^{-1} \mathbf{X} ~\text{, and}~\left(\tilde{\mathbf{X}}_{rw}^{(item)}\right)^{\top} = \mathbf{D}_I^{-1} \mathbf{X}^{\top} .
\end{equation}

Using RW normalization, the original gram matrix ${\mathbf{P}} = \mathbf{X^{\top}} \mathbf{X}$ is converted to $\tilde{\mathbf{P}}_{rw}= \mathbf{D}_I^{-1} \mathbf{X^{\top}} \mathbf{D}_U^{-1} \mathbf{X}$. To analyze the effect of item-side normalization, if we assume that $\mathbf{D}_U = \mathbf{I}$, the normalized gram matrix is represented by $\tilde{\mathbf{P}}_{rw}^{(item)} = \mathbf{D}_I^{-1} \mathbf{X^{\top}} \mathbf{X}$. Since $\mathbf{D}_I^{-1}$ performs row-wise item normalization in $\mathbf{X^{\top}} \mathbf{X}$, each entry $(\tilde{\mathbf{P}}_{rw}^{(item)})_{ij}$ indicates the co-occurrence of two items $i$ and $j$ normalized by the popularity for the source item $i$.
\begin{equation} \label{eq:rw_item_normalization}
    \left(\tilde{\mathbf{P}}_{rw}^{(item)}\right)_{ij} = \left(\mathbf{D}_I^{-1}{\mathbf{X^{\top}} \mathbf{X}}\right)_{ij} = \left(\mathbf{D}_I^{-1}{\mathbf{P}}\right)_{ij} = \frac{freq(i,j)}{freq(i)}.
\end{equation}

Although RW normalization mitigates the popularity of source items, it does not consider the popularity of target items.

\vspace{1mm}
\noindent
\textbf{Sym normalization}.
Unlike RW normalization, it normalizes the input matrix $\mathbf{X}$ in both user and item sides.
\begin{equation} \label{eq:x_sym_norm}
    \tilde{\mathbf{X}}_{sym} = \mathbf{D}_U^{-1/2} \mathbf{X} \mathbf{D}_I^{-1/2}.
\end{equation}

Using Sym normalization, the gram matrix is updated into $\tilde{\mathbf{P}}_{sym} = \mathbf{D}_I^{-1/2} \mathbf{X^{\top}}\mathbf{D}_U^{-1} \mathbf{X} \mathbf{D}_I^{-1/2}$. Assuming that $\mathbf{D}_U = \mathbf{I}$, the gram matrix equals $\tilde{\mathbf{P}}_{sym}^{(item)} = \mathbf{D}_I^{-1/2} \mathbf{X^{\top}} \mathbf{X} \mathbf{D}_I^{-1/2}$. Because it deals with both row- and column-wise item normalization, $(\tilde{\mathbf{P}}_{sym}^{(item)})_{ij}$ is normalized by the popularity of both the source item $i$ and the target item $j$.
\begin{align} \label{eq:sym_item_normalization}
    \left(\tilde{\mathbf{P}}_{sym}^{(item)}\right)_{ij} = \left(\mathbf{D}_I^{-1/2}{\mathbf{P}}\mathbf{D}_I^{-1/2}\right)_{ij} = \frac{freq(i,j)}{freq(i)^{1/2}freq(j)^{1/2}}.
\end{align}

In contrast to RW normalization, Sym normalization can penalize the popularity of target items, making it effective in directly alleviating the recommendation of popular items. However, it employs the same normalization weight for source and target items, despite their different influences in the gram matrix $\tilde{\mathbf{P}}_{sym}$.

\section{Normalized Linear Autoencoders}\label{sec:analysis}

This section presents the incorporation of user/item normalization in the objective function of LAEs, focusing on RW and Sym normalization methods and their limitations.

\subsection{Generalized Normalization to LAEs}

To deal with user/item normalization, we formulate a generalized objective function of LAEs using a diagonal user weight matrix\footnote{We assume all the weight matrices as the diagonal matrices, not the full matrices. We leave them as a future design choice.} $\mathbf{W}_U \in \mathbb{R}^{m \times m}$ and two diagonal item weight matrices $\mathbf{W}_I^{(1)}, \mathbf{W}_I^{(2)} \in \mathbb{R}^{n \times n}$. Note that we can extend the objective function to other LAEs with additional constraints, such as EASE$^\text{R}$~\cite{Steck19EASE} and RLAE~\cite{MoonKL23RDLAE}.
\begin{equation} \label{eq:user_item_obj}
    \min_{\mathbf{B}} \| \mathbf{W}_U (\mathbf{X}\mathbf{W}_I^{(1)} - \mathbf{X}\mathbf{W}_I^{(1)}\mathbf{B}) \|_F^2 + \lambda \|\mathbf{W}_I^{(2)} \mathbf{B}\|_F^2.
\end{equation}

For user normalization, the user weight matrix $\mathbf{W}_U$ modulates user importance in the reconstruction error, where setting $\mathbf{W}_U = \mathbf{D}_U$ adjusts the influence of active users.
For item normalization, $\mathbf{W}_I^{(1)}$ transforms the gram matrix into $\mathbf{W}_{I}^{(1)}\mathbf{X}^{\top} \mathbf{X}\mathbf{W}_{I}^{(1)}$, enabling row- and column-wise weighting that controls both source and target item weights in computing $\mathbf{B}$.
The weight matrix $\mathbf{W}_I^{(2)}$ provides row-wise weighting in the L2 regularization term, controlling the weight of source items in computing $\mathbf{B}$.
By adopting item normalization $\mathbf{D}_{I}$ for $\mathbf{W}_I^{(1)}$ and $\mathbf{W}_I^{(2)}$, we can mitigate item popularity bias by weakening the importance of popular items.

Through simple convex optimization of Eq.~\eqref{eq:user_item_obj}, we readily derive the following closed-form solution.
\begin{align} \label{eq:user_item_solution}
\resizebox{1\hsize}{!}{$
    \hat{\mathbf{B}}_{gen} = \left(\tilde{\mathbf{P}}_{gen}+\lambda\mathbf{I}\right)^{-1} \tilde{\mathbf{P}}_{gen},~\text{where}~\tilde{\mathbf{P}}_{gen} = (\mathbf{W}_I^{(2)})^{-2}\mathbf{W}_I^{(1)} \mathbf{X}^{\top}\mathbf{W}^{2}_U\mathbf{X}\mathbf{W}_I^{(1)}.
$}
\end{align}

User normalization is performed in between the input matrix $\mathbf{X}$ (\ie, $\mathbf{X}^{\top}\mathbf{W}^{2}_U\mathbf{X}$). For item normalization, row-wise (\ie, $(\mathbf{W}_I^{(2)})^{-2}\mathbf{W}_I^{(1)}$) and column-wise (\ie, $\mathbf{W}_I^{(1)}$) normalization is used before and after the gram matrix, respectively. This reveals that $\mathbf{W}_I^{(1)}$ solely cannot differentiate source/target item normalization, necessitating $\mathbf{W}_I^{(2)}$.

\subsection{Existing Normalization for LAEs}

\vspace{1mm}
\noindent
\textbf{RW normalization}. It performs user normalization and row-wise item normalization. We thus replace $\mathbf{W}_U$ and $\mathbf{W}_I^{(2)}$ in Eq.~\eqref{eq:user_item_obj} with $\mathbf{D}_U^{-1/2}$ and $\mathbf{D}_I^{1/2}$, respectively.
\begin{equation} \label{eq:rw_objective}
    \min_{\mathbf{B}} \|\mathbf{D}_U^{-1/2} (\mathbf{X}-\mathbf{X}\mathbf{B})\|_F^2 + \lambda \|\mathbf{D}_I^{1/2} \mathbf{B}\|_F^2.
\end{equation}

The same solution form is derived by substituting the original gram matrix $\mathbf{P}$ with the RW normalized gram matrix $\tilde{\mathbf{P}}_{rw}$.
\begin{equation}
    \hat{\mathbf{B}}_{rw} = \left(\tilde{\mathbf{P}}_{rw} + \lambda \mathbf{I}\right)^{-1}\left(\tilde{\mathbf{P}}_{rw}\right),~\text{where}~\tilde{\mathbf{P}}_{rw} = \mathbf{D}_I^{-1} \mathbf{X^{\top}} \mathbf{D}_U^{-1} \mathbf{X}. \label{eq:rw_solution}
\end{equation}

\vspace{1mm}
\noindent
\textbf{Sym normalization}. Unlike RW normalization, it handles both row- and column-wise item normalization. For $\mathbf{W}_U$ and $\mathbf{W}_I^{(1)}$, we adopt two diagonal degree matrices $\mathbf{D}_U^{-1/2}$ and $\mathbf{D}_I^{-1/2}$, respectively.
\begin{equation} \label{eq:sym_objective}
    \min_{\mathbf{B}} \| \mathbf{D}_U^{-1/2} (\mathbf{X}\mathbf{D}_I^{-1/2} - \mathbf{X}\mathbf{D}_I^{-1/2}\mathbf{B}) \|_F^2 + \lambda \|\mathbf{B}\|_F^2.
\end{equation}

We also derive the solution by only replacing the gram matrix $\mathbf{P}$ with the Sym normalized gram matrix $\tilde{\mathbf{P}}_{sym}$.
\begin{equation} 
\resizebox{1\hsize}{!}{$
    \hat{\mathbf{B}}_{sym} = \left(\tilde{\mathbf{P}}_{sym} + \lambda \mathbf{I}\right)^{-1}\left(\tilde{\mathbf{P}}_{sym}\right),~\text{where}~\tilde{\mathbf{P}}_{sym} = \mathbf{D}_I^{-1/2} \mathbf{X^{\top}}\mathbf{D}_U^{-1} \mathbf{X} \mathbf{D}_I^{-1/2}. \label{eq:sym_solution} 
$}
\end{equation}

Using the generalized objective function in Eq.~\eqref{eq:user_item_obj}, we discuss the limitations of existing normalization methods. Firstly, RW normalization only considers source item popularity, while Sym normalization considers the source and target items with equal weight. Besides, both of them assign the equal weight to the user and item sides (\eg, For Sym normalization, $\mathbf{D}_U^{-1/2}$ and $\mathbf{D}_I^{-1/2}$). To address these limitations, we (i) fully utilize three weight matrices (\ie, $\mathbf{W}_U$, $\mathbf{W}_I^{(1)}$, and $\mathbf{W}_I^{(2)}$) and (ii) adaptively modulate the popularity for items and users by considering the dataset characteristics.

\section{Data-Adaptive Normalization (DAN)}\label{sec:model}

This section proposes a simple yet effective normalization method called \textbf{DAN}. According to the dataset characteristics, it adjusts the importance of items and users through its key components, \ie, \emph{item- and user-adaptive normalization}.

The objective function of LAEs using DAN employs all three normalization weight matrices in Eq.~\eqref{eq:user_item_obj}.
\begin{align} \label{eq:dan_objective}
    \min_{\mathbf{B}} \|\mathbf{D}_U^{-\beta/2} (\mathbf{X}\mathbf{D}_I^{-\alpha}-\mathbf{X}\mathbf{D}_I^{-\alpha}\mathbf{B})\|_F^2 + \lambda \|\mathbf{D}_I^{1/2-\alpha}\mathbf{B}\|_F^2.
\end{align}

Depending on $\alpha$ and $\beta$, the objective function is equivalent to that of RW normalized LAEs in Eq.~\eqref{eq:rw_objective} for $\alpha=0$ and $\beta=1$, or to that of Sym normalized LAEs in Eq.~\eqref{eq:sym_objective} for $\alpha=1/2$ and $\beta=1$. The closed-form solution of LAEs with DAN is as follows.
\begin{align} \label{eq:dan_solution}
\resizebox{1\hsize}{!}{$
    \hat{\mathbf{B}}_{dan} = \left(\Tilde{\mathbf{P}}_{dan} + \lambda \mathbf{I}\right)^{-1}\left(\Tilde{\mathbf{P}}_{dan}\right), \text{where}~\Tilde{\mathbf{P}}_{dan} = \mathbf{D}_I^{-(1-\alpha)} \mathbf{X^{\top}} \mathbf{D}_U^{-\beta} \mathbf{X} \mathbf{D}_I^{-\alpha}.
$}
\end{align}

To thoroughly analyze the effect of DAN, we dissect the DAN gram matrix $\Tilde{\mathbf{P}}_{dan}$ in Eq.~\eqref{eq:dan_solution} into item and user aspects.

\subsection{Item-Adaptive Normalization}
The first component of DAN uses the parameter $\alpha$ to adjust popularity bias, whose properties we prove in Theorem~\ref{theorem:item_adap_norm}. (The detailed proof can be found in Appendix~\ref{appen:proof_item_adap_norm}.)

\begin{theorem} \label{theorem:item_adap_norm}
    Item-adaptive normalization (i) provides a denoising effect~\cite{Steck20edlae} and (ii) controls popularity bias: A larger $\alpha$ alleviates target items' popularity bias, while a smaller $\alpha$ focuses on source items' popularity bias.
    \begin{align}
        \hat{\mathbf{B}}_{LAE} (\mathbf{P}=\mathbf{D}_I^{-(1-\alpha)} \mathbf{X^{\top}} \mathbf{X} \mathbf{D}_I^{-\alpha}) = 
        \mathbf{D}_{I}^{\alpha} \hat{\mathbf{B}}_{DLAE} \mathbf{D}_{I}^{-\alpha}. \label{eq:item_norm_final}
    \end{align}
\end{theorem}

Here, $\hat{\mathbf{B}}_{DLAE}$ means the closed-form solution of DLAE~\cite{Steck20edlae}, which applies a dropout to the input matrix and reconstructs the original matrix (\ie, the denoising process)\footnote{Appendix~\ref{appen:robustness} shows DAN's superior robustness against noisy inputs.}. Then, $\mathbf{D}_{I}^{-\alpha}$ mitigates the item popularity bias by penalizing the target items' popularity on the right-hand side of the weight matrix $\hat{\mathbf{B}}_{DLAE}$. A higher $\alpha$ penalizes target items' popularity more strongly, particularly effective for datasets with low Gini index~\cite{ChinCC22} where users interact with various items evenly.

\begin{figure}

\centering
\begin{tabular}{cc}
\renewcommand{\arraystretch}{1} 
\includegraphics[height=2.55cm]{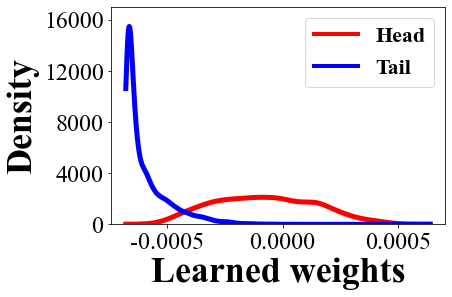} &
\includegraphics[height=2.55cm]{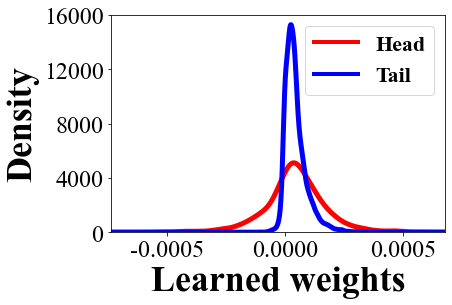} \\
\multicolumn{1}{c}{(a) W/O normalization} &\multicolumn{1}{c}{(b) $\alpha=0$} \vspace{2mm} \\
\includegraphics[height=2.55cm]{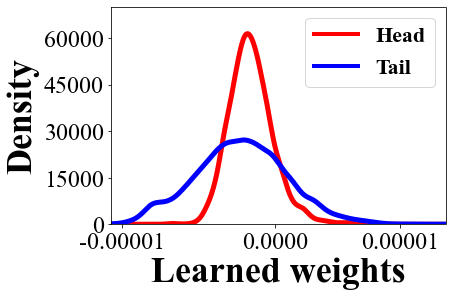} &
\includegraphics[height=2.55cm]{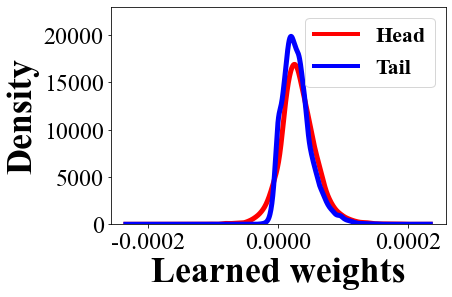} \\
\multicolumn{1}{c}{(c) $\alpha=0.5$} &\multicolumn{1}{c}{(d) $\alpha=0.2$ (ours)}

\vspace{-2mm}
\end{tabular}

\caption{Distribution of the weights learned~\cite{Steck19EASE} with different $\alpha$ values on ML-20M. The red and blue lines are the estimated probability density functions (PDFs) for head and tail items. The weights are averaged over $\hat{\mathbf{B}}$ in a column-wise direction. The x-axis is the average weight of items, and the area under the curve corresponds to the probability of having the weights within that range.}\label{fig:ml-20m_distribution}

\vspace{-3mm}
\end{figure}

To empirically validate Theorem~\ref{theorem:item_adap_norm}, we analyze the weight distributions of $\hat{\mathbf{B}}$ for head and tail items under different $\alpha$. 
Figure~\ref{fig:ml-20m_distribution} illustrates the probability density functions, where weight magnitude determines an item's recommendation likelihood. If normalization is not applied (Figure~\ref{fig:ml-20m_distribution}(a)), head items dominate with large weights due to their frequent occurrences. When $\alpha=0$ (\ie, $\mathbf{D}_I^{-1} \mathbf{X^{\top}} \mathbf{X}$), the normalization partially mitigates the head item dominance by considering only source item popularity. When $\alpha=0.5$ (\ie, $\mathbf{D}_I^{-1/2} \mathbf{X^{\top}} \mathbf{X} \mathbf{D}_I^{-1/2}$), normalization of source and target items further reduces the head-tail distribution gap, but large differences remain.
In contrast, item-adaptive normalization ($\alpha=0.2$) achieves the most balanced distribution through dynamic adjustment.

Beyond benefits, item-adaptive normalization preserves eigenvalues.
In Eq.~\eqref{eq:item_norm_final}, since $\mathbf{D}_{I}^{\alpha} \hat{\mathbf{B}}_{DLAE} \mathbf{D}_{I}^{-\alpha}$ is a similarity transformation of $\hat{\mathbf{B}}_{DLAE}$, they share identical eigenvalues for any $\alpha \in \mathbb{R}$, ensuring neighborhood bias remains invariant.

\subsection{User-Adaptive Normalization}

User-adaptive normalization controls neighborhood bias using the parameter $\beta$.
Excluding item-adaptive normalization (\ie, $\mathbf{D}_I^{-(1-\alpha)}$ and $\mathbf{D}_I^{-\alpha}$) from Eq.~\eqref{eq:dan_solution}, the LAE solution with user-adaptive normalization is as follows. 
\begin{align} 
    \hat{\mathbf{B}}_{LAE} (\mathbf{P}= & \mathbf{X^{\top}} \mathbf{D}_U^{-\beta} \mathbf{X})  = \left(\mathbf{X}^\top \mathbf{D}_U^{-\beta} \mathbf{X} + \lambda \mathbf{I}\right)^{-1} \mathbf{X}^\top \mathbf{D}_U^{-\beta} \mathbf{X}. \label{eq:user_norm_solution}
\end{align}

Through eigen-decomposition, the matrix $\mathbf{\tilde{P}}_{user} = \mathbf{X^{\top}} \mathbf{D}_U^{-\beta} \mathbf{X}$ is decomposed into three matrices. The eigenvalues of $\mathbf{\tilde{P}}_{user}$ follow the descending order, \ie, $1 \geq \mu_1 \geq \cdots \geq \mu_n \geq 0$.
\begin{equation}\label{eq:eigendecomp}
    \tilde{\mathbf{P}}_{user} = \mathbf{V} \text{diag}(\mu_1, \mu_2, \dots, \mu_n) \mathbf{V}^{\top} = \sum_{i=1}^n \mu_i \mathbf{v}_i \mathbf{v}_i^{\top},
\end{equation}
where $\mathbf{V} \in \mathbb{R}^{n \times n}$ represents the eigenvectors, and $\mathbf{v}_i \in \mathbb{R}^{n}$ is an $i$-th eigenvector of the matrix $\tilde{\mathbf{P}}_{user}$.
From the perspective of signal processing~\cite{OrtegaFKMV18}, high eigenvalues (\eg, $\mu_1$ and $\mu_2$) correspond to low-frequency signals capturing global item relationships, while low eigenvalues (\eg, $\mu_{n-1}$ and $\mu_n$) represent high-frequency components encoding local neighborhood correlations, which are predominantly formed by active users. Neighborhood bias can be mitigated by emphasizing low-frequency components while suppressing high-frequency ones (\ie, higher $\beta$). Notably, normalization efficiently modulates eigenvalues without costly eigen-decomposition~\cite{YangSLQZY22}.

Theorem~\ref{theorem:user_adap_norm} characterizes user-adaptive normalization building on Lemmas~\ref{lemma:weight_gram_eigenvalues} and \ref{lemma:rayleigh_eigenvalues}. (Full proof is presented in Appendix~\ref{appen:proof_user_adap_norm}.)

\begin{lemma}[Eigenvalue Relationship between Weight Matrix and Gram Matrix] \label{lemma:weight_gram_eigenvalues}
Following previous work~\cite{MoonKL23RDLAE}, for weight matrix $\hat{\mathbf{B}} = (\mathbf{{P}} + \lambda \mathbf{I})^{-1} \mathbf{{P}}$, its eigenvalues $\gamma_i$ can be expressed in terms of the eigenvalues $\mu_i$ of gram matrix $\mathbf{{P}}$ as:
\begin{equation}
    \gamma_i = \frac{\mu_i}{\mu_i + \lambda}, \quad \text{for all } i = 1, 2, \dots, n.
\end{equation}
\end{lemma}

\begin{lemma}[Monotonicity of Eigenvalues via Rayleigh Quotient] \label{lemma:rayleigh_eigenvalues}
    For a symmetric matrix $\mathbf{A} \in \mathbb{R}^{n \times n}$, if the Rayleigh quotient $R( \mathbf{A}, \mathbf{v}) = \frac{\mathbf{v}^{\top} \mathbf{A} \mathbf{v}}{\mathbf{v}^{\top} \mathbf{v}}$ decreases for all non-zero vectors $\mathbf{v} \in \mathbb{R}^n$, then all eigenvalues of $\mathbf{A}$ decrease.
\end{lemma}

\begin{theorem} \label{theorem:user_adap_norm}
    User-adaptive normalization adjusts the eigenvalues ($\gamma_1, \dots, \gamma_n$) of the weight matrix $\mathbf{B}$: A larger $\beta$ pushes eigenvalues toward zero, while a smaller $\beta$ keeps them closer to one. For any $\beta_1 > \beta_2 \geq 0$, all eigenvalues strictly decrease, i.e., $\gamma_i(\beta_1) < \gamma_i(\beta_2)$ for all $i$.
\end{theorem}

\begin{figure}
\centering
\begin{tabular}{cc}
\includegraphics[height=2.9cm]{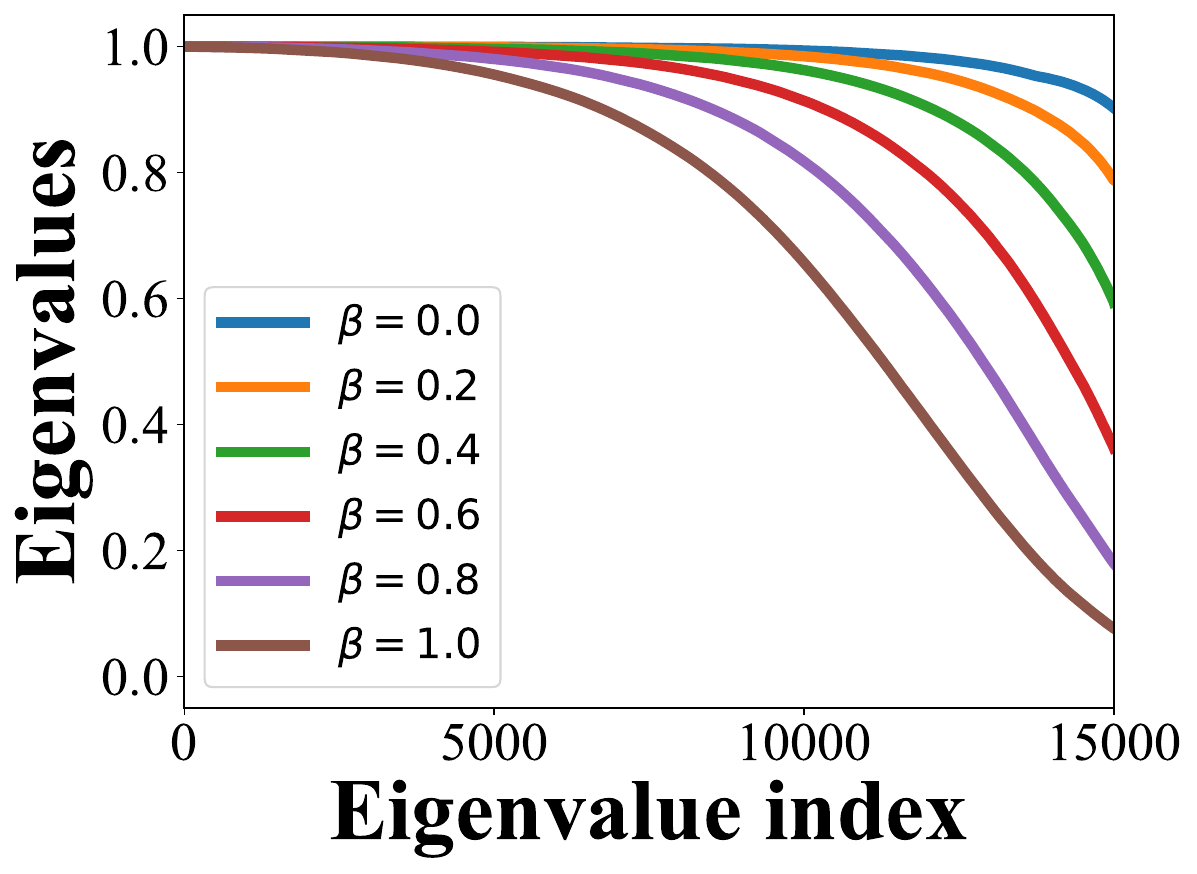} &
\includegraphics[height=2.9cm]{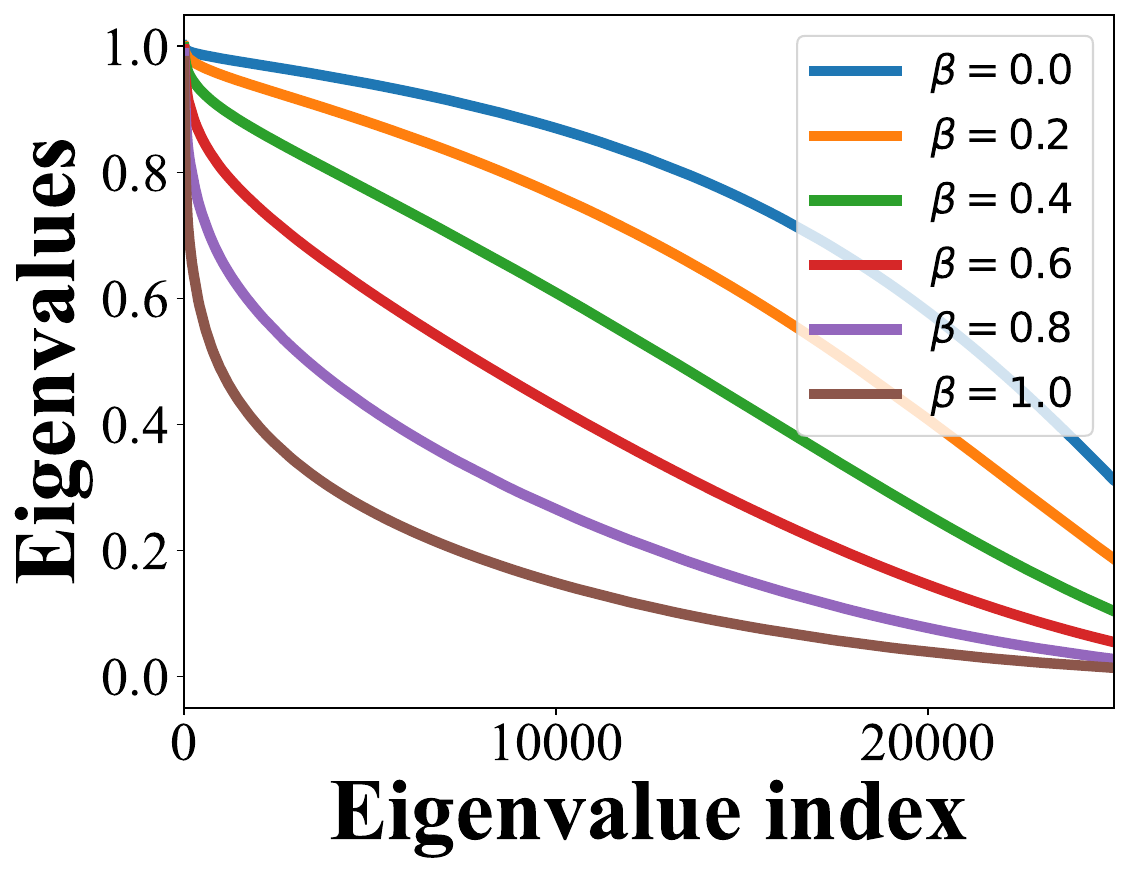} \\
\multicolumn{1}{c}{(a) ML-20M} &\multicolumn{1}{c}{(b) Yelp2018}

\end{tabular}

\vspace{-2mm}
\caption{Eigenvalue distribution of the weight matrix $\mathbf{\hat{B}}$ according to $\beta$ on ML-20M and Yelp2018.}\label{fig:eigenvalue_dist}
\vspace{-3mm}
\end{figure}

To empirically verify Theorem~\ref{theorem:user_adap_norm}, we observe the eigenvalue distribution depending on $\beta$. Figure~\ref{fig:eigenvalue_dist} shows the eigenvalue distributions of two datasets, with high-frequency components increasing from left to right.
When $\beta$ increases from $0$ to $1$, most eigenvalues gradually decrease and approach zero, while a few low-frequency eigenvalues remain significant. This shows that user-adaptive normalization can balance low- and high-frequency signals while preserving essential global patterns, thus adjusting neighborhood bias.

The prior study~\cite{YanCC0DYT23} demonstrates that high-homophilic datasets primarily leverage low-frequency information rather than high-frequency components. While the recent study~\cite{JiangG0CY24} has proposed using Jaccard similarity to measure homophily in recommender systems, we introduce a more nuanced \emph{weighted homophily ratio}:
\begin{equation}\label{eq:our_homophily}
    \mathcal{H}_{w} = \frac{ \sum_{(i,j) \in \mathcal{E}} w_{ij} \cdot s_{ij}}{\sum_{(i,j) \in \mathcal{E}} w_{ij}},~\text{where}~w_{ij} = {|\mathcal{V}_i \cap \mathcal{V}_j|^{\delta}} \cdot \frac{{|\mathcal{V}_i \cap \mathcal{V}_j|}}{min({|\mathcal{V}_i|, |\mathcal{V}_j|})}
\end{equation}

Here, $s_{ij}= \frac{|\mathcal{V}_i \cap \mathcal{V}_i|}{|\mathcal{V}_i \cup \mathcal{V}_j|} \in [0,1]$ is Jaccard similarity between user sets for items $i$ and $j$, and $\mathcal{V}_i$ is a set of users who interacted with item $i$.
A high homophily ratio means that items are largely connected with homogeneous items, implying low neighborhood bias.

The weighted homophily ratio has two improvements:
\begin{itemize}[leftmargin=5mm]
    \item \textbf{Absolute intersection size:}  
    The first term $|\mathcal{V}_i \cap \mathcal{V}_j|^{\delta}$ increases as the number of shared users grows.  
    For example, given two item pairs with $s_{ij} = \frac{100}{200}$ and $\frac{10}{20}$, the absolute count of 100 suggests a more relevant relationship than 10. We set $\delta$ as 1.5.
    
    \item \textbf{Subset relationship:}  
    The second term $\frac{|\mathcal{V}_i \cap \mathcal{V}_j|}{\min(|\mathcal{V}_i|, |\mathcal{V}_j|)}$ increases when one item set is nearly a subset of another.  
    It reflects relatedness when a smaller set is entirely contained within a larger one, also known as the \emph{Szymkiewicz-Simpson coefficient}~\cite{VermaA20Szymkiewicz}.
\end{itemize}

Based on the weighted homophily ratio, user-adaptive normalization adjusts $\beta$ to control neighborhood bias. Highly engaged users tend to form local item relationships through numerous interactions, which can dominate and obscure global patterns. For high-homophilic datasets where similar items naturally form dense connections, a higher $\beta$ helps emphasize these global patterns by reducing the influence of individual user interactions. Conversely, for low-homophilic datasets where items exhibit more diverse relationships, a lower $\beta$ preserves the local neighborhood information captured in individual user preferences.

\begin{table}[t] \small
\begin{center}
\caption{Statistics of six benchmark datasets. $Gini_{i}$ denotes the Gini index~\cite{ChinCC22} that measures the skewness of item distributions. $\mathcal{H}_{w}$ means the weighted homophily ratio, in Eq.~\eqref{eq:our_homophily}.}
\vspace{-2mm}
\label{tab:statistics}
\begin{tabular}{P{1.65cm}|P{0.7cm}P{0.7cm}|P{0.7cm}P{0.9cm}|P{0.6cm}P{0.6cm}}
\toprule
Dataset     & \#Users   & \#Items   & \#Inter.     & Density   & $Gini_{i}$ & $\mathcal{H}_{w}$ \\
\midrule
ML-20M      & 136,677   & 20,108    & 10.0M         & 0.36\%    & 0.90 &  0.109  \\
Netflix     & 463,435   & 17,769    & 56.9M         & 0.69\%    & 0.86 &  0.127  \\
MSD         & 571,355   & 41,140    & 33.6M         & 0.36\%    & 0.56 &  0.123  \\
Gowalla     & 29,858    & 40,981    & 1.03M     & 0.08\%    & 0.44 &  0.085 \\
Yelp2018    & 31,668    & 38,048    & 1.56M     & 0.13\%    & 0.51 &  0.044 \\
Amazon-book & 52,643    & 91,599    & 2.98M     & 0.06\%    & 0.46 &  0.059 \\
\bottomrule
\end{tabular}
\end{center}
\vspace{-2mm}
\end{table}

\section{Experimental Setup}\label{sec:setup}

\noindent
\textbf{Datasets}. We conduct experiments on six benchmark datasets, \ie, ML-20M, Netflix, MSD, Gowalla, Yelp2018, and Amazon-book, as summarized in Table~\ref{tab:statistics}. Following~\cite{LiangKHJ18MultVAE, ShenWZSZLL21GFCF, ChoiHPC23BSPM, MoonKL23RDLAE}, for ML-20M and Netflix, we used a 5-core setting, and for MSD, we used only users with at least 20 songs and songs listened to by at least 200 users. For Gowalla, Yelp2018, and Amazon-book, we used a 10-core setting.

\vspace{1mm}
\noindent
\textbf{Evaluation protocols}. We evaluate DAN using two protocols: (i) \emph{Strong generalization} evaluates users who were not seen in the training phase, so we split the training, validation, and test sets into 8:1:1 based on the user side~\cite{MoonKL23RDLAE, Steck19EASE}. In the inference phase, we assume that users in the validation and test sets have 80\% of their total ratings and evaluate for the remaining 20\%. 
(ii) \emph{Weak generalization} only considers users covered in the training phase. Following the convention~\cite{0001DWLZ020LightGCN, ChoiHPC23BSPM, MoonKL23RDLAE, ShenWZSZLL21GFCF}, we split the ratings of the entire interaction matrix into 8:2 to serve as training and test sets.

\vspace{1mm}
\noindent
\textbf{Evaluation metrics}. We adopt two ranking metrics, \ie, \emph{Recall} and \emph{NDCG}. To validate the overall performance and the effect of mitigating popularity bias, we introduce three different measures: \textit{Average-over-all (AOA)}, \textit{Tail}, and \textit{Unbiased}~\cite{YangCXWBE18Unbias, LeePLL22BISER, MoonKL23RDLAE} evaluation. We split items into head (top 20\% popular) and tail (bottom 80\% unpopular) groups based on popularity. AOA is the measure for both item groups, while Tail is measured for only tail items. For unbiased evaluation, we set $\gamma=2$ along~\cite{LeePLL22BISER, YangCXWBE18Unbias} to ensure that the effect of popularity bias is not reflected in the evaluation.

\vspace{1mm}
\noindent
\textbf{Baseline models}. We equip DAN to three linear autoencoder models, \ie, LAE, EASE$^\text{R}$~\cite{Steck19EASE}, and RLAE~\cite{MoonKL23RDLAE}. We used DLAE, EDLAE~\cite{Steck20edlae}, and RDLAE~\cite{MoonKL23RDLAE}, which add a denoising effect to the previous three models. Since item-adaptive normalization integrates the denoising effect, we skip experiments on the denoising LAEs with DAN. We also employed eight CF models as the baselines. MultVAE~\cite{LiangKHJ18MultVAE}, LightGCN~\cite{0001DWLZ020LightGCN}, and XSimGCL~\cite{YuXCCHY24XSimGCL} are representative non-linear models. GF-CF~\cite{ShenWZSZLL21GFCF}, BSPM~\cite{ChoiHPC23BSPM}, Turbo-CF~\cite{ParkSS2024turbocf}, SVD-AE~\cite{HongCLKP24svdae}, and SGFCF~\cite{PengLSM24sgfcf} are representative linear models.
Among these, only five models (MultVAE, GF-CF, BSPM, Turbo-CF, and SVD-AE) can be evaluated under strong generalization. For brevity, we denote EASE$^\text{R}$ as EASE.

\vspace{1mm}
\noindent
\textbf{Reproducibility}. All models are implemented in the same framework code as~\cite{ShenWZSZLL21GFCF, ChoiHPC23BSPM, MoonKL23RDLAE}. For LAEs without DAN, $\lambda$ was searched in [10, 20, ..., 500, 1000], and with DAN in [1e-3, 2e-3, ..., 20, 50]. For the denoising LAEs (\ie, DLAE, EDLAE, and RDLAE), the dropout ratio $p$ was searched in [0.1, 0.9] with step size 0.1. $\alpha$ and $\beta$ were searched in [0, 0.5] and [0, 1] with step size 0.1, respectively.
For weak generalization protocol, we use the reported results of GF-CF, BSPM, MultVAE, and LightGCN from~\cite{ChoiHPC23BSPM}. For all other models, we reproduced the experiments by referring to the best hyperparameters in their original papers.
Since linear models are deterministic, we conducted a single run, while the performance of non-linear models was averaged over five seeds. We run all experiments on NVIDIA RTX-3090 24GB GPU and Intel Xeon Gold 6226R CPU.

\begin{table} \small 
\caption{Performance comparison on ML-20M, Netflix, and MSD under the strong generalization. `LAE$_\text{DAN}$' indicates LAE equipped with our proposed DAN. The best results are marked in \red{bold}, and the second best in \blue{underlined}.}
\label{tab:strong_result_top3data}
\vspace{-2mm}
\begin{center}
\renewcommand{\arraystretch}{1} 
\begin{tabular}{P{0.3cm}|P{1.1cm}|P{0.7cm}P{0.7cm}|P{0.7cm}P{0.7cm}|P{0.7cm}P{0.7cm}}
\toprule
\multirow{2}{*}{\rotatebox{90}{}} & \multirow{2}{*}{Model}  & \multicolumn{2}{c|}{AOA} & \multicolumn{2}{c|}{Tail}  & \multicolumn{2}{c}{Unbiased} \\
\multirow{17}{*}{\rotatebox{90}{ML-20M}} &  & R@20 & N@20 & R@20 & N@20 & R@20 & N@20 \\
\midrule
& MultVAE & 0.3895 & 0.3278 & 0.0133 & 0.0077 & \red{0.2996} & \red{0.0509} \\
& GF-CF & 0.3250 & 0.2736 & 0.0029 & 0.0022 & 0.2188 & 0.0371 \\ 
& BSPM & 0.3725 & 0.3183 & 0.0104 & 0.0060 & 0.2761 & 0.0463 \\
& TurboCF & 0.3276 & 0.2728 & 0.0166 & 0.0111 & 0.2336 & 0.0402 \\
& SVD-AE & 0.3720 & 0.3205 & 0.0075 & 0.0041 & 0.2672 & 0.0454 \\ 
\cmidrule{2-8}
& LAE & 0.3757 & 0.3228 & 0.0005 & 0.0001 & 0.2827 & 0.0473 \\
& EASE & 0.3905 & 0.3390 & 0.0052 & 0.0022 & 0.2857 & 0.0479 \\
& RLAE & 0.3913 & 0.3402 & 0.0137 & 0.0069 & {0.2951} & 0.0487 \\
& DLAE & 0.3923 & 0.3408 & 0.0084 & 0.0047 & 0.2898 & 0.0477 \\ 
& EDLAE & 0.3925 & 0.3421 & 0.0066 & 0.0035 & 0.2859 & 0.0480 \\ 
& RDLAE & {0.3932} & {0.3422} & {0.0123} & {0.0062} & \blue{0.2987} & {0.0489} \\
\cmidrule{2-8}
& LAE$_\text{DAN}$ & 0.3930 & 0.3414 & \red{0.0234} & \red{0.0128} & 0.2925 & {0.0493} \\ 
& EASE$_\text{DAN}$ & \blue{0.3950} & \blue{0.3430} & {0.0217} & {0.0120} & {0.2955} & {0.0497} \\
& RLAE$_\text{DAN}$ & \red{0.3956} & \red{0.3432} & \blue{0.0227} & \blue{0.0127} & {0.2973} & \blue{0.0501} \\
\midrule
\multirow{14}{*}{\rotatebox{90}{Netflix}}
& MultVAE & 0.3434 & 0.3129 & 0.0291 & 0.0215 & 0.2432 & 0.0345 \\
& GF-CF & 0.2972 & 0.2724 & 0.0185 & 0.0123 & 0.1868 & 0.0264 \\ 
& BSPM & 0.3163 & 0.2909 & 0.0339 & 0.0227 & 0.2145 & 0.0302 \\
& TurboCF & 0.2826 & 0.2586 & 0.0265 & 0.0188 & 0.1777 & 0.0279 \\
& SVD-AE & 0.3206 & 0.2977 & 0.0289 & 0.0186 & 0.2121 & 0.0308 \\ 
\cmidrule{2-8}
& LAE & 0.3465 & 0.3237 & 0.0066 & 0.0036 & 0.2357 & 0.0326 \\
& EASE & 0.3618 & 0.3388 & 0.0404 & 0.0222 & 0.2554 & 0.0351 \\
& RLAE & {0.3623} & {0.3392} & {0.0585} & {0.0377} & {0.2606} & {0.0355} \\
& DLAE & 0.3621 & 0.3400 & 0.0597 & 0.0381 & 0.2549 & 0.0355 \\ 
& EDLAE & 0.3659 & 0.3428 & 0.0470 & 0.0279 & 0.2569 & 0.0358 \\  
& RDLAE & {0.3661} & {0.3431} & 0.0545 & 0.0344 & {0.2598} & {0.0360} \\
\cmidrule{2-8}
& LAE$_\text{DAN}$ & {0.3631} & {0.3405} & {0.0623} & \blue{0.0411} & {0.2598} & {0.0363} \\
& EASE$_\text{DAN}$ & \red{0.3666} & \blue{0.3433} & \red{0.0658} & \red{0.0424} & \red{0.2646} & \red{0.0371} \\ 
& RLAE$_\text{DAN}$ & \blue{0.3662} & \red{0.3434} & \blue{0.0628} & {0.0400} & \blue{0.2628} & \blue{0.0370} \\
\midrule
\multirow{14}{*}{\rotatebox{90}{MSD}}
& MultVAE & 0.2443 & 0.2270 & 0.1372 & 0.0988 & 0.2013 & 0.0258 \\
& GF-CF & 0.2513 & 0.2457 & 0.1727 & 0.1331 & 0.2137 & 0.0282 \\ 
& BSPM & 0.2682 & 0.2616 & 0.2121 & 0.1583 & 0.2494 & 0.0321 \\
& TurboCF & 0.2666 & 0.2593 & 0.2153 & 0.1652 & 0.2497 & 0.0324 \\
& SVD-AE & 0.2859 & 0.2743 & 0.1984 & 0.1379 & 0.2502 & 0.0305 \\ 
\cmidrule{2-8}
& LAE & 0.2848 & 0.2740 & 0.1862 & 0.1234 & 0.2568 & 0.0320 \\
& EASE & {0.3338} & \blue{0.3261} & 0.2504 & 0.1758 & 0.3019 & 0.0377 \\
& RLAE & {0.3338} & \blue{0.3261} & {0.2507} & {0.1767} & {0.3021} & {0.0378} \\
& DLAE & 0.3288 & 0.3208 & {0.2526} & {0.1863} & 0.2993 & 0.0378 \\ 
& EDLAE & {0.3336} & 0.3258 & 0.2503 & 0.1782 & 0.3014 & 0.0378 \\   
& RDLAE & \blue{0.3341} & \red{0.3265} & 0.2511 & 0.1784 & {0.3022} & {0.0379} \\
\cmidrule{2-8}
& LAE$_\text{DAN}$ & {0.3290} & {0.3209} & {0.2530} & \blue{0.1873} & {0.2999} & {0.0380} \\
& EASE$_\text{DAN}$ & {0.3336} & {0.3259} & \red{0.2621} & \red{0.1926} & \red{0.3071} & \red{0.0389} \\
& RLAE$_\text{DAN}$ & \red{0.3342} & \red{0.3265} & \blue{0.2573} & {0.1864} & \blue{0.3049} & \blue{0.0384} \\
\bottomrule
\end{tabular}
\vspace{-2mm}
\end{center}
\end{table}

\begin{table} \small 
\caption{Performance comparison on Gowalla, Yelp2018, and Amazon-book under the strong generalization. `LAE$_\text{DAN}$' indicates LAE equipped with our proposed DAN. The best results are marked in \red{bold}, and the second best in \blue{underlined}.} 
\label{tab:strong_result_bot3data}
\vspace{-2mm}
\begin{center}
\renewcommand{\arraystretch}{1} 
\begin{tabular}{P{0.3cm}|P{1.1cm}|P{0.7cm}P{0.7cm}|P{0.7cm}P{0.7cm}|P{0.7cm}P{0.7cm}}
\toprule
\multirow{2}{*}{\rotatebox{90}{}} & \multirow{2}{*}{Model}  & \multicolumn{2}{c|}{AOA} & \multicolumn{2}{c|}{Tail}  & \multicolumn{2}{c}{Unbiased} \\
\multirow{17}{*}{\rotatebox{90}{Gowalla}} &  & R@20 & N@20 & R@20 & N@20 & R@20 & N@20 \\ \midrule
& MultVAE & 0.1788 & 0.1269 & 0.0698 & 0.0381 & 0.1289 & 0.0256 \\
& GF-CF & 0.2252 & 0.1660 & 0.1151 & 0.0591 & 0.1734 & 0.0343 \\ 
& BSPM & 0.2373 & 0.1757 & 0.1270 & 0.0638 & 0.1849 & 0.0360 \\
& TurboCF & 0.2281 & 0.1686 & 0.1166 & 0.0627 & 0.1770 & 0.0357 \\
& SVD-AE & 0.2292 & 0.1717 & 0.0898 & 0.0414 & 0.1639 & 0.0317 \\ 
\cmidrule{2-8}
& LAE & 0.2271 & 0.1706 & 0.0799 & 0.0371 & 0.1672 & 0.0326 \\
& EASE & 0.2414 & 0.1831 & 0.0941 & 0.0428 & 0.1753 & 0.0335 \\    
& RLAE & {0.2448} & {0.1873} & {0.1243} & {0.0625} & {0.1912} & {0.0370} \\
& DLAE & {0.2495} & {0.1891} & 0.1109 & 0.0532 & 0.1881 & 0.0366 \\
& EDLAE & 0.2469 & 0.1859 & 0.0951 & 0.0432 & 0.1790 & 0.0344 \\  
& RDLAE & {0.2499} & {0.1900} & {0.1210} & {0.0587} & {0.1923} & {0.0373} \\ 
\cmidrule{2-8}
& LAE$_\text{DAN}$ & 0.2491 & 0.1911 & \red{0.1392} & \red{0.0741} & \red{0.2003} & \red{0.0398} \\
& EASE$_\text{DAN}$ & \red{0.2527} & \blue{0.1918} & {0.1306} & {0.0670} & {0.1983} & {0.0392} \\  
& RLAE$_\text{DAN}$ & \blue{0.2520} & \red{0.1919} & \blue{0.1332} & \blue{0.0694} & \blue{0.1988} & \blue{0.0393} \\
\midrule
\multirow{14}{*}{\rotatebox{90}{Yelp2018}}
& MultVAE & 0.0963 & 0.0747 & 0.0193 & 0.0121 & 0.0643 & 0.0074 \\
& GF-CF & 0.1134 & 0.0900 & 0.0155 & 0.0078 & 0.0685 & 0.0081 \\ 
& BSPM & 0.1198 & 0.0951 & 0.0251 & 0.0129 & 0.0776 & 0.0090 \\
& TurboCF & 0.1144 & 0.0930 & 0.0218 & 0.0133 & 0.0735 & 0.0092 \\
& SVD-AE & 0.1145 & 0.0923 & 0.0147 & 0.0076 & 0.0693 & 0.0082 \\ 
\cmidrule{2-8}
& LAE & 0.1160 & 0.0954 & 0.0086 & 0.0039 & 0.0705 & 0.0086 \\
& EASE & 0.1144 & 0.0933 & 0.0091 & 0.0042 & 0.0679 & 0.0081 \\
& RLAE & {0.1173} & {0.0968} & 0.0127 & 0.0060 & {0.0735} & {0.0089} \\
& DLAE & 0.1190 & 0.0971 & 0.0121 & 0.0057 & 0.0724 & 0.0087 \\
& EDLAE & 0.1171 & 0.0957 & 0.0103 & 0.0049 & 0.0698 & 0.0084 \\   
& RDLAE & {0.1190} & {0.0976} & {0.0161} & {0.0077} & {0.0741} & {0.0089} \\
\cmidrule{2-8}
& LAE$_\text{DAN}$ & {0.1230} & {0.1002} & \red{0.0313} & \red{0.0175} & \red{0.0834} & \blue{0.0100} \\ 
& EASE$_\text{DAN}$ & \red{0.1238} & \red{0.1011} & {0.0294} & {0.0163} & {0.0828} & \blue{0.0100} \\ 
& RLAE$_\text{DAN}$ & \blue{0.1237} & \blue{0.1010} & \blue{0.0302} & \blue{0.0169} & \blue{0.0832} & \red{0.0101} \\
\midrule 
\multirow{14}{*}{\rotatebox{90}{Amazon-book}}
& MultVAE & 0.1005 & 0.0816 & 0.0374 & 0.0250 & 0.0771 & 0.0096 \\
& GF-CF & 0.1668 & 0.1492 & 0.0988 & 0.0702 & 0.1401 & 0.0195 \\ 
& BSPM & 0.1742 & 0.1569 & 0.1156 & 0.0819 & 0.1546 & 0.0212 \\
& TurboCF & 0.1720 & 0.1557 & 0.0957 & 0.0662 & 0.1454 & 0.0199 \\
& SVD-AE & 0.1433 & 0.1239 & 0.0579 & 0.0372 & 0.1065 & 0.0139 \\ 
\cmidrule{2-8}
& LAE & 0.1920 & 0.1749 & 0.1012 & 0.0635 & 0.1644 & 0.0220 \\
& EASE & 0.1912 & 0.1734 & 0.0761 & 0.0444 & 0.1481 & 0.0195 \\
& RLAE & {0.1968} & {0.1804} & {0.1057} & 0.0672 & {0.1649} & {0.0221} \\
& DLAE & {0.1994} & {0.1820} & 0.0993 & 0.0631 & 0.1637 & 0.0220 \\
& EDLAE & 0.1940 & 0.1756 & 0.0829 & 0.0512 & 0.1523 & 0.0205 \\
& RDLAE & {0.2011} & {0.1834} & {0.1043} & {0.0670} & {0.1663} & {0.0225} \\
\cmidrule{2-8}
& LAE$_\text{DAN}$ & 0.1979 & 0.1811 & \red{0.1314} & \red{0.0886} & \red{0.1766} & \red{0.0239} \\
& EASE$_\text{DAN}$ & \blue{0.2017} & \blue{0.1835} & {0.1136} & {0.0746} & {0.1705} & {0.0231} \\ 
& RLAE$_\text{DAN}$ & \red{0.2019} & \red{0.1836} & \blue{0.1226} & \blue{0.0820} & \blue{0.1747} & \blue{0.0236} \\
\bottomrule
\end{tabular}
\vspace{-2mm}
\end{center}
\end{table}

\begin{table}\small

\caption{Performance comparison on Gowalla, Yelp2018, and Amazon-book with the weak generalization. `LAE$_\text{DAN}$' indicates LAE equipped with our proposed DAN. The best results are marked in \red{bold}, and the second best in \blue{underlined}.}
\vspace{-2mm}
\label{tab:weak_result}
\begin{center}
\renewcommand{\arraystretch}{1}
\begin{tabular}{P{1.7cm}|P{0.7cm}P{0.7cm}|P{0.7cm}P{0.7cm}|P{0.7cm}P{0.7cm}}
\toprule
Dataset & \multicolumn{2}{c|}{Gowalla} & \multicolumn{2}{c|}{Yelp2018} & \multicolumn{2}{c}{Amazon-book} \\
Model & R@20 & N@20 & R@20 & N@20 & R@20 & N@20 \\
\midrule
MultVAE~\cite{LiangKHJ18MultVAE} & 0.1641 & 0.1335 & 0.0584 & 0.0450 & 0.0407 & 0.0315 \\
LightGCN~\cite{0001DWLZ020LightGCN} & 0.1830 & 0.1554 & 0.0649 & 0.0530 & 0.0411 & 0.0315 \\
XSimGCL~\cite{YuXCCHY24XSimGCL} & 0.1861 & 0.1580 & 0.0711 & 0.0584 & 0.0541 & 0.0420 \\
\midrule
GF-CF~\cite{ShenWZSZLL21GFCF} & 0.1849 & 0.1536 & 0.0697 & 0.0571 & 0.0710 & 0.0584 \\
BSPM~\cite{ChoiHPC23BSPM} & \blue{0.1920} & \blue{0.1597} & \red{0.0720} & \red{0.0593} & 0.0733 & 0.0609 \\
TurboCF~\cite{ParkSS2024turbocf} & {0.1835} & 0.1531 & 0.0693 & {0.0574} & 0.0730 & 0.0611 \\
SVD-AE~\cite{HongCLKP24svdae} & {0.1860} & 0.1550 & 0.0683 & {0.0571} & 0.0569 & 0.0451 \\
SGFCF~\cite{PengLSM24sgfcf} & {0.1899} & 0.1566 & \blue{0.0713} & \blue{0.0588} & 0.0694 & 0.0565 \\
\midrule
LAE & 0.1630 & 0.1295 & 0.0658 & 0.0555 & 0.0746 & 0.0611 \\
EASE~\cite{Steck19EASE} & 0.1765 & 0.1467 & 0.0657 & 0.0552 & 0.0710 & 0.0566 \\
RLAE~\cite{MoonKL23RDLAE} & 0.1772 & 0.1467 & 0.0667 & 0.0562 & {0.0754} & {0.0615} \\
DLAE~\cite{Steck20edlae} & 0.1839 & 0.1533 & 0.0678 & 0.0570 & 0.0751 & 0.0610 \\
EDLAE~\cite{Steck20edlae} & 0.1844 & 0.1539 & 0.0673 & 0.0565 & 0.0711 & 0.0566 \\
RDLAE~\cite{MoonKL23RDLAE} & 0.1845 & 0.1539 & 0.0679 & 0.0569 & 0.0754 & {0.0613} \\
\midrule
LAE$_\text{DAN}$ & 0.1901 & 0.1591 & 0.0703 & 0.0586 & \blue{0.0759} & \blue{0.0627} \\
EASE$_\text{DAN}$ & 0.1905 & 0.1594 & {0.0706} & 0.0587 & \red{0.0762} & \red{0.0630} \\
RLAE$_\text{DAN}$ & \red{0.1922} & \red{0.1605} & {0.0706} & 0.0587 & \red{0.0762} & \red{0.0630} \\
\bottomrule
\end{tabular}
\vspace{-2mm}
\end{center}
\end{table}

\section{Experimental Results}\label{sec:results}


\subsection{Performance Comparison}\label{sec:overall_performance}

\noindent
\textbf{Strong generalization}. Tables~\ref{tab:strong_result_top3data}--\ref{tab:strong_result_bot3data} present the performance of state-of-the-art CF models in non-LAE-based and LAE-based groups. We found three observations: (i) LAE equipped with DAN performs better on high-homophilic datasets. On high-homophilic datasets (ML-20M, Netflix, MSD), LAE$_{DAN}$ shows an average AOA performance gain of 9.36\% at NDCG@20 compared to LAE, while on low-homophilic datasets (Gowalla, Yelp2018, Amazon-book), it shows a 6.86\% gain. This demonstrates that normalization effectively captures global correlations in datasets with more global patterns. (ii) DAN consistently achieves the best performance in tail and unbiased evaluation, mitigating popularity bias across backbone models and datasets. On ML-20M and Yelp2018, RLAE$_\text{DAN}$ results in Tail performance gain of 84.06\% and 181.67\% at NDCG@20, respectively.
(iii) Most LAEs with DAN outperform the existing denoising LAE models (DLAE, EDLAE, RDLAE), thanks to DAN's inherent denoising effect.

\vspace{1mm}
\noindent
\textbf{Weak generalization}. Table~\ref{tab:weak_result} presents the performance of state-of-the-art CF models in three groups: non-linear models, linear models (excluding LAEs), and LAE-based models. (i) Despite weak generalization, DAN consistently enhances LAE model performance. LAEs equipped with DAN surpass both the baseline LAEs (\ie, LAE, EASE, and RLAE) and their denoising versions, with RLAE$_\text{DAN}$ demonstrating superior performance among DAN-equipped models. (ii) RLAE$_\text{DAN}$ achieves better or comparable performance to state-of-the-art linear model BSPM~\cite{ChoiHPC23BSPM}, outperforming BSPM on Amazon-book by up to 3.45\% at NDCG@20. In particular, RLAE$_\text{DAN}$ performs well on sparse datasets such as Gowalla and Amazon-book. (iii) Furthermore, all LAEs with DAN outperform non-linear models, with RLAE$_\text{DAN}$ notably achieving up to 50\% performance gains over XSimGCL~\cite{YuXCCHY24XSimGCL} at NDCG@20 on Amazon-book.

To validate the generalizability of DAN, we applied DAN to a linear model SLIST~\cite{ChoiKLSL21} proposed for the session-based recommendation. We found that DAN significantly improves the performance of the linear session-based recommender model. (Refer to Appendix~\ref{appen:session_comparison})

\begin{figure}
\begin{tabular}{cc}
\includegraphics[height=2.7cm]{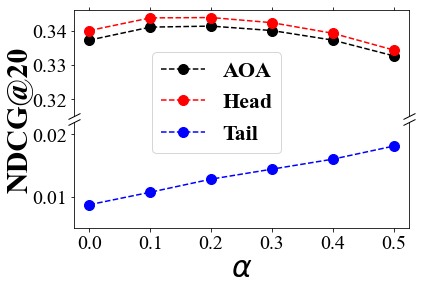} &
\includegraphics[height=2.7cm]{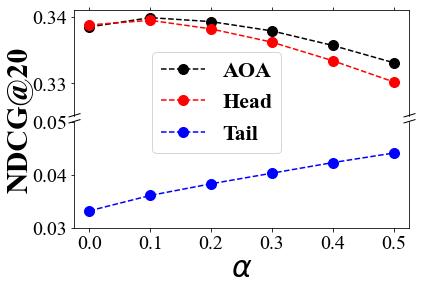} \\
\multicolumn{1}{c}{(a) ML-20M} &\multicolumn{1}{c}{(b) Netflix} \vspace{2mm} \\
\includegraphics[height=2.7cm]{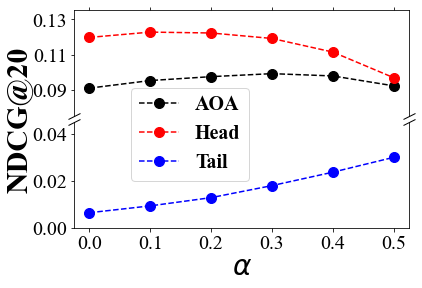} &
\includegraphics[height=2.7cm]{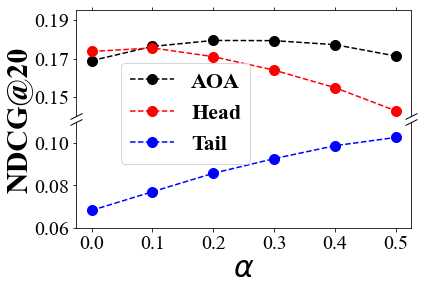} \\

\multicolumn{1}{c}{(c) Yelp2018} &\multicolumn{1}{c}{(d) Amazon-book}
\vspace{-2mm}
\end{tabular}
\caption{NDCG@20 of LAE$_\text{DAN}$ over $\alpha$ for item normalization on four datasets. When adjusting $\alpha$, we keep $\beta$ fixed at the optimal parameter. }\label{fig:alpha}
%
\vspace{-2mm}
\end{figure}
\begin{figure}
\begin{tabular}{cc}
\includegraphics[height=2.7cm]{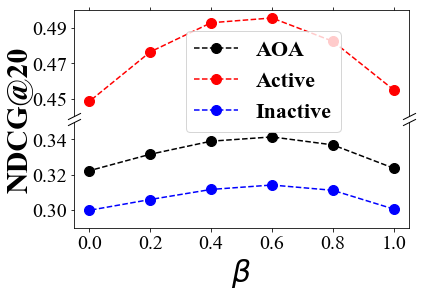} &
\includegraphics[height=2.7cm]{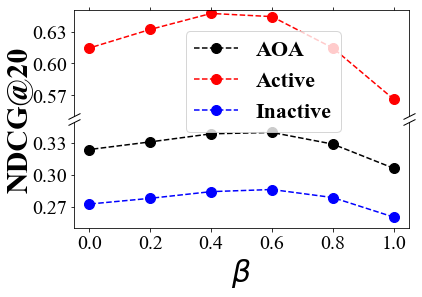} \\
\multicolumn{1}{c}{(a) ML-20M} &\multicolumn{1}{c}{(b) Netflix} \vspace{2mm} \\
\includegraphics[height=2.7cm]{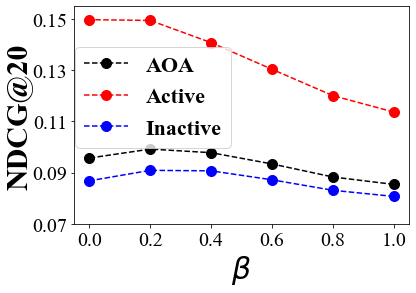} &
\includegraphics[height=2.7cm]{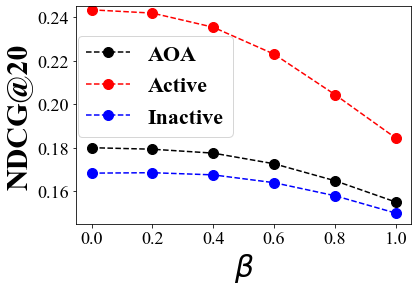} \\
\multicolumn{1}{c}{(c) Yelp2018} &\multicolumn{1}{c}{(d) Amazon-book}
\vspace{-2mm}
\end{tabular}
\caption{NDCG@20 of LAE$_\text{DAN}$ over $\beta$ for user normalization on four datasets. Active and Inactive mean the performance of the top 20\% users with high-activity and the rest. When adjusting $\beta$, we keep $\alpha$ fixed at the optimal parameter. }\label{fig:inactive_user}
\vspace{-2mm}
\end{figure}

\subsection{Hyperparameter Sensitivity}\label{sec:hyperparameter}

\noindent
\textbf{Effect of item normalization}. Figure~\ref{fig:alpha} illustrates the head and tail item performance, over adjusting $\alpha$.
For brevity, Figures~\ref{fig:alpha}-\ref{fig:inactive_user} and Table~\ref{tab:dataset_results_two} report only NDCG@20; Recall@20 showed similar trends.
Appendix~\ref{appen:sensitivity} shows the results in the remaining datasets, \ie, MSD and Gowalla.
\begin{itemize}[leftmargin=5mm]
    \item On the four datasets, Tail performance increases while Head performance decreases as $\alpha$ increase. This is due to the higher $\alpha$ strengthening the normalization for the target items. It directly mitigates the popularity bias in the recommendation process and makes unpopular items more likely to be recommended.

    \vspace{0.5mm}
    \item To understand the importance of adjusting $\alpha$, we compare the difference between the best and worst AOA performance for each dataset: ML-20M and Netflix have a difference of 2.62\% and 2.51\%, respectively; Yelp2018 and Amazon-book have a difference of 9.00\% and 6.76\%, respectively. The relatively lower performance difference of DAN on ML-20M and Netflix compared to Yelp2018 and Amazon-book is attributed to the larger $Gini_{i}$ of ML-20M and Netflix. (For statistics on $Gini_{i}$, refer to Table~\ref{tab:statistics}.) It means that recommending popular items may have the AOA performance advantage in ML-20M and Netflix. Thus, \uline{we suggest adjusting $\alpha$ lower for datasets with high $Gini_{i}$ and higher for datasets with low $Gini_{i}$.}
\end{itemize}

\vspace{1mm}
\noindent
\textbf{Effect of user normalization}.
Figure~\ref{fig:inactive_user} depicts the performance for two user groups (\ie, Active and Inactive) depending on $\beta$.
\begin{itemize}[leftmargin=5mm]
    \item For high-homophilic datasets, \ie, ML-20M and Netflix, both Active and Inactive performances trend up and down as $\beta$ increases. This is because these datasets have global interaction patterns favoring popular items. Specifically, ML-20M and Netflix have large $\mathcal{H}_{w}$ of 0.109 and 0.127, respectively. By moderately penalizing active users, DAN can improve the performance of both user groups because it learns global item correlations enough from those of inactive users while discarding unnecessary patterns (\eg, interaction noises) from active users. However, if we over-penalize active users, such as RW and Sym normalization, both user groups have a performance drop.

    \vspace{0.5mm}
    \item For low-homophilic datasets, \ie, Yelp2018 and Amazon-book, both Active and Inactive performances consistently decrease as $\beta$ increases. This is due to the diverse local patterns in these datasets; $\mathcal{H}_{w}$ for Yelp2018 and Amazon-book are 0.044 and 0.059, respectively. Thus, penalizing active users degrades both performances because it is difficult to learn the diverse high-frequency patterns from those of inactive users alone. To summarize, \uline{we recommend tuning $\beta$ higher for datasets with high $\mathcal{H}_{w}$ and lower for datasets with low $\mathcal{H}_{w}$.}
\end{itemize}

\begin{table} \small

\caption{Performance comparison for various normalization methods on ML-20M and Yelp2018. The backbone model is the LAE, and the metric is NDCG@20. Appendix~\ref{appen:norm_comparison} provides the results in the remaining datasets.}
\vspace{-2mm}
\label{tab:dataset_results_two}
\begin{center}
\renewcommand{\arraystretch}{1} 
\begin{tabular}{P{1.5cm}|P{2.0cm}|P{0.9cm}P{0.9cm}P{0.9cm}}

\toprule
Dataset   & Method      & AOA         & Head         & Tail                 \\
\midrule
\multirow{6}{*}{ML-20M}
& W/O norm       & 0.3228       & 0.3264       & 0.0001    \\
& RW norm       & {0.3358}       & {0.3386}       & 0.0088       \\ 
& Sym norm       & 0.3321       & 0.3338       & \red{0.0184}      \\
& User norm      & \blue{0.3390}       & \blue{0.3428}       & 0.0001      \\
& Item norm      & {0.3341}       & 0.3379       & \blue{0.0140}      \\
& DAN (ours)     & \red{0.3414}       & \red{0.3433}       & {0.0128}         \\
\midrule
\multirow{6}{*}{Yelp2018}
& W/O norm       & 0.0954       & \red{0.1285}       & 0.0039        \\
& RW norm      & 0.0840       & 0.1128       & 0.0044          \\ 
& Sym norm     & 0.0922       & 0.1052       & \blue{0.0217}             \\
& User norm      & \blue{0.0966}       & \blue{0.1274}       & 0.0115      \\
& Item norm      & 0.0965       & 0.1225       & \red{0.0249}      \\
& DAN (ours)    & \red{0.1002}       & 0.1212       & 0.0175          \\
\bottomrule
\end{tabular}
\vspace{-2mm}

\end{center}
\end{table}

\subsection{Comparing Various Normalization} \label{sec:ablation_study}
Table~\ref{tab:dataset_results_two} reports the performance of applying different normalization methods to LAE on the strong generalization.
We also conducted a case study for these normalization methods in Appendix~\ref{appen:case_study}.
\begin{itemize}[leftmargin=5mm]
    \item On both datasets, DAN demonstrates the highest AOA performance while significantly improving Tail performance, meaning that it can adaptively normalize items and users. For ML-20M, RW and Sym normalizations improve AOA, Head, and Tail performance compared to LAE without normalization (\ie, W/O). In contrast, for Yelp2018, both normalizations decreased AOA performance but improved Tail performance compared to W/O. This reveals the limitations of RW and Sym normalizations, which cannot normalize depending on the dataset.

    \vspace{0.5mm}
    \item Item-adaptive normalization (\ie, Item norm) consistently improves Tail performance. For ML-20M and Yelp2018, Tail performance gain is 13,900\% and 538\%, respectively. On the other hand, user-adaptive normalization (\ie, User norm) shows different tendencies over the datasets. For ML-20M, it has a gain of 5.02\% in Head performance only, while for Yelp2018, it has a gain of 195\% in Tail performance only. This is because users in ML-20M strongly prefer popular items, while users in Yelp2018 prefer popular and unpopular items similarly.
            
\end{itemize}

\begin{table} \small

\caption{Efficiency comparison for DAN and SOTA models on three datasets. Train and Infer indicate runtime (in seconds) for training and inference. The inference time is measured with a batch size of 4,096 for all test users.}

\vspace{-2mm}
\label{tab:efficiency}
\begin{center}
\renewcommand{\arraystretch}{1} 
\begin{tabular}{P{1.7cm}|r r|r r|r r}

\toprule
Dataset         & \multicolumn{2}{c|}{Gowalla}          & \multicolumn{2}{c|}{Yelp2018}         & \multicolumn{2}{c}{Amazon-book} \\
Model & Train & Infer & Train & Infer & Train & Infer \\
\midrule
XSimGCL~\cite{YuXCCHY24XSimGCL}           & 967 & 6 & 1,420 & 6 & 9,587 & 18 \\  
BSPM~\cite{ChoiHPC23BSPM}               & 74 & 257 & 80 & 180 & 43 & 3,211 \\ 
LAE                                     & 132 & 28 & 94 & 26 & 1,279 & 215 \\ 
LAE$_\text{DAN}$                        & 139 & 28 & 98 & 26 & 1,346 & 215 \\ 

\bottomrule
\end{tabular}
\vspace{-2mm}

\end{center}
\end{table}

\subsection{Efficiency} \label{sec:efficiency}
Table~\ref{tab:efficiency} compares the computational costs between DAN and state-of-the-art models (\ie, XSimGCL~\cite{YuXCCHY24XSimGCL} and BSPM~\cite{ChoiHPC23BSPM})\footnote{All models were evaluated on CPU for fair comparison, except XSimGCL~\cite{YuXCCHY24XSimGCL} which required GPU for reasonable runtime.}. (i) LAE demonstrates superior speed in both training and inference compared to other models. On Gowalla, LAE achieves 608\% and 198\% faster execution times than XSimGCL and BSPM, respectively, in total training and inference time. (ii) The integration of DAN introduces minimal overhead, with LAE$_\text{DAN}$ showing only 5.30\%, 4.26\%, and 5.24\% increases in training time across the three datasets. This additional cost is attributed to the multiplication of $\mathbf{D}_U$ and $\mathbf{D}_I$ by the gram matrix $\mathbf{X}^{\top}\mathbf{X}$ during training. The efficiency of DAN enables rapid hyperparameter optimization, particularly as LAE$_\text{DAN}$ requires only three parameters (\ie, $\lambda$, $\alpha$, and $\beta$), while eight and seven parameters for XSimGCL and BSPM, respectively.

\section{Related Work}

\vspace{1mm}
\noindent
\textbf{Linear recommender models}. They are classified into latent factor-based models and neighborhood-based models. While latent factor-based models decompose the user-item interaction matrix into lower-dimensional latent factors, neighborhood-based models typically use a co-occurrence matrix as an item-item similarity matrix. SLIM~\cite{NingK11SLIM} is an item neighborhood-based model that learns with the objective function of ridge regression. EASE$^\text{R}$~\cite{Steck19EASE} derives a closed-form solution using L2-regularization and zero-diagonal constraints from SLIM~\cite{NingK11SLIM}. DLAE/EDLAE~\cite{Steck20edlae} learns with stronger regularization for popular items, while RLAE/RDLAE~\cite{MoonKL23RDLAE} relieves diagonal constraints to emphasize unpopular items. Unlike RLAE, which indirectly adjusts the effect of users/items via the weight matrix $\mathbf{B}$, our work directly modulates the user-item matrix $\mathbf{X}$. Furthermore, HIGHER~\cite{SteckL21Higher} leverages high-order connectivity, where three or more items appear in one user simultaneously. Recently, SVD-AE~\cite{HongCLKP24svdae} utilizes truncated SVD on the reconstructed matrix to improve robustness against noise in the user-item matrix.
SGFCF~\cite{PengLSM24sgfcf} applies graph filtering that is generalizable to noise removal tailored to dataset density variations.

\vspace{1mm}
\noindent
\textbf{Normalization}. In recommender systems, RW~\cite{CooperLRS14P3alpha, PaudelCNB17RP3beta} and Sym normalization methods~\cite{Wang0WFC19NGCF, 0001DWLZ020LightGCN, FuPLW22, ChoiHPC23BSPM} are widely used to relieve popularity bias~\cite{0007D0F0023, Zhu0ZZWC21} by suppressing the effects of active users and popular items. Some works~\cite{CooperLRS14P3alpha, PaudelCNB17RP3beta} using RW normalization prevent the transition probability to popular item nodes in a user-item bipartite graph from becoming too large. Also, Sym normalization~\cite{0001DWLZ020LightGCN, Wang0WFC19NGCF, YuXCCHY24XSimGCL, ChoiHPC23BSPM} is used in graph convolution networks (GCN) to prevent some nodes from propagating dominantly throughout the graph. \citet{abs-1904-13033} presents the only work in LAEs that makes use of normalization by heuristically applying column-wise weighting to the closed-form solution of EASE$^\text{R}$~\cite{Steck19EASE}\footnote{Appendix~\ref{appen:col_item_norm} shows that the relation between DAN and the normalization~\cite{abs-1904-13033}.}. Also,~\citet{WLCZDWSLW22} performs normalization on the adjacency matrix during the neighborhood aggregation process of GCN-based models.
Recently, SGFCF~\cite{PengLSM24sgfcf} applies normalization to reduce high-frequency signals for denoising but lacks distinction between user and item normalization in its popularity bias analysis.
Meanwhile, a few studies~\cite{Chen23, KimPK23, GuptaGMVS2019} have shown the effectiveness of normalization in latent space by resizing the length of user/item embeddings. In contrast, we focus on normalization suitable for efficient linear models without embeddings.

\section{Conclusion}\label{sec:conclusion}

This paper proposed a simple yet effective normalization, named  \emph{\textbf{D}ata-\textbf{A}daptive \textbf{N}ormalization (\textbf{DAN})}. The existing LAEs with DAN demonstrated up to 128.57\% and 12.36\% performance gains for long-tail items and unbiased evaluation scenarios across six benchmark datasets. Additionally, DAN can be easily tailored to reflect the characteristics of the datasets, such as the skewness of the distribution over items (\ie, $Gini_{i}$) and the homophily ratio (\ie, $\mathcal{H}_{w}$). We observed that RW and Sym normalization methods alleviate popularity bias while they utilize fixed weights between source/target items and users. Unlike these normalizations, DAN can dynamically adjust source/target item popularity and neighborhood biases. 
DAN can be extended to neural models by normalizing the input user-item matrix or intermediate layers as future work.

\begin{acks}
    This work was partly supported by the Institute of Information \& communications Technology Planning \& evaluation (IITP) grant and the National Research Foundation of Korea (NRF) grant funded by the Korea government (MSIT) (No. RS-2019-II190421, IITP-2025-RS-2020-II201821, RS-2022-II221045, IITP-2025-RS-2024-00437633, and RS-2025-00564083, each contributing 20\% to this research).
\end{acks}


\appendix

\section{Theoretical Proofs} \label{appen:convex_optimization}

\subsection{Item-Adaptive Normalization} \label{appen:proof_item_adap_norm}
The LAE solution with item-adaptive normalized gram matrix $\mathbf{D}_I^{-(1-\alpha)} \mathbf{X^{\top}} \mathbf{X} \mathbf{D}_I^{-\alpha}$ can be expanded as follows.
\begin{align}
    & \hat{\mathbf{B}}_{LAE} (\mathbf{P}=\mathbf{D}_I^{-(1-\alpha)} \mathbf{X^{\top}} \mathbf{X} \mathbf{D}_I^{-\alpha}) \nonumber \\
    = & \left(\mathbf{D}_I^{-(1-\alpha)} \mathbf{X^{\top}} \mathbf{X} \mathbf{D}_I^{-\alpha}  + \lambda \mathbf{I}\right)^{-1} \mathbf{D}_I^{-(1-\alpha)} \mathbf{X^{\top}} \mathbf{X} \mathbf{D}_I^{-\alpha} \nonumber \\
    = & \left(\mathbf{D}_I^{-(1-\alpha)} (\mathbf{X^{\top}} \mathbf{X} + \lambda \mathbf{D}_I) \mathbf{D}_I^{-\alpha}\right)^{-1} \mathbf{D}_I^{-(1-\alpha)} \mathbf{X^{\top}} \mathbf{X} \mathbf{D}_I^{-\alpha} \nonumber \\ 
    = &~\mathbf{D}_I^{\alpha} \left(\mathbf{X^{\top}} \mathbf{X} + \lambda \mathbf{D}_I\right)^{-1} \mathbf{D}_I^{1-\alpha} \mathbf{D}_I^{-(1-\alpha)} \mathbf{X^{\top}} \mathbf{X} \mathbf{D}_I^{-\alpha} \nonumber  \\
    = &~\mathbf{D}_I^{\alpha} \left(\mathbf{X^{\top}} \mathbf{X} + \lambda \mathbf{D}_I\right)^{-1} \mathbf{X^{\top}} \mathbf{X} \mathbf{D}_I^{-\alpha} =~\mathbf{D}_{I}^{\alpha} \hat{\mathbf{B}}_{DLAE} \mathbf{D}_{I}^{-\alpha} \label{eq:item_norm_final_proof}
\end{align}

In Eq.~\eqref{eq:item_norm_final_proof}, $\left(\mathbf{X^{\top}} \mathbf{X} + \lambda \mathbf{D}_I\right)^{-1} \mathbf{X^{\top}} \mathbf{X}$ is identical to the closed-form solution $\hat{\mathbf{B}}_{DLAE}$ of DLAE~\cite{Steck20edlae} with dropout probability $p=\frac{\lambda}{1+\lambda}$.\footnote{A study~\cite{Steck20edlae} found that applying dropout to LAEs is equivalent to performing a weighted regularization by item popularity. $\hat{\mathbf{B}}_{DLAE} = \left(\mathbf{P} + \boldsymbol{\Lambda}\right)^{-1}\mathbf{P}$, where $\boldsymbol{\Lambda} = \frac{p}{1-p} \cdot \text{diagMat} \left(\text{diag} (\mathbf{P}) \right)$. Since $\mathbf{P}_{i,i}$ means the popularity of item $i$, it is straightforward to see that $\text{diagMat} \left(\text{diag} (\mathbf{P}) \right)$ is equal to $\mathbf{D}_I$.}

\subsection{User-Adaptive Normalization} \label{appen:proof_user_adap_norm}

For any non-zero vector $\mathbf{v} \in \mathbb{R}^n$, the Rayleigh quotient of $\tilde{\mathbf{P}}_{user} (\beta)=\mathbf{X}^{\top} \mathbf{D}_U^{-\beta} \mathbf{X}$ is:
\begin{align}
\resizebox{1\hsize}{!}{$
    R(\tilde{\mathbf{P}}_{user} (\beta), \mathbf{v}) = \frac{\mathbf{v}^{\top} \tilde{\mathbf{P}}_{user} (\beta) \mathbf{v}}{\mathbf{v}^{\top} \mathbf{v}} 
               = \frac{(\mathbf{X} \mathbf{v})^{\top} \mathbf{D}_U^{-\beta} (\mathbf{X}\mathbf{v})} {\mathbf{v}^{\top} \mathbf{v}}
               = \frac{\sum_{k=1}^m (\mathbf{X}\mathbf{v})_k^2 d_k^{-\beta}}{\mathbf{v}^{\top} \mathbf{v}} \nonumber
$}
\end{align}
where $(\mathbf{X}\mathbf{v})_k$ is the weighted sum of item interactions for user $k$. Since $d_k > 1$ for all users $k$, for $\beta_1 > \beta_2$: $d_k^{-\beta_1} < d_k^{-\beta_2}$.
Therefore,
\begin{align}
\resizebox{1\hsize}{!}{$
R(\tilde{\mathbf{P}}_{user} (\beta_1), \mathbf{v}) = \frac{\sum_{k=1}^m (\mathbf{X}\mathbf{v})_k^2 d_k^{-\beta_1}}{\mathbf{v}^{\top} \mathbf{v}} 
              < \frac{\sum_{k=1}^m (\mathbf{X}\mathbf{v})_k^2 d_k^{-\beta_2}}{\mathbf{v}^{\top} \mathbf{v}}
              = R(\tilde{\mathbf{P}}_{user} (\beta_2), \mathbf{v}) \nonumber
$}
\end{align}

By Lemma~\ref{lemma:rayleigh_eigenvalues}, since this inequality holds for all non-zero vectors $\mathbf{v}$ and $\tilde{P}_{user}$ is symmetric, we can conclude that:
\begin{equation}
    \mu_i(\beta_1) < \mu_i(\beta_2) \text{ for all } i
\end{equation}

By Lemma~\ref{lemma:weight_gram_eigenvalues}, since $\gamma_i(\beta) = \frac{\mu_i(\beta)}{\mu_i(\beta) + \lambda}$ is strictly increasing in $\mu_i(\beta)$, we can conclude that $\gamma_i(\beta_1) < \gamma_i(\beta_2)$ for all $i$. \qedhere

\subsection{Relation between DAN and Column-wise Item Normalization~\cite{abs-1904-13033}} \label{appen:col_item_norm}
We theoretically show that DAN includes column-wise item normalization~\cite{abs-1904-13033}. The LAE solution for column-wise item normalization unfolds as follows.
\begin{align}
    & \frac{\partial}{\partial \mathbf{B}}(\min _{\mathbf{B}}\|(\mathbf{X} \mathbf{D}_I^{-\gamma}-\mathbf{X B})\|_F^2+\lambda\|\mathbf{B}\|_F^2)=0 \\
    \Leftrightarrow & -2 \mathbf{X}^{\top}(\mathbf{X} \mathbf{D}_I^{-\gamma}-\mathbf{X B})+2 \lambda \mathbf{B}=0 \\
    \therefore &~\hat{\mathbf{B}}=\left(\mathbf{X}^{\top} \mathbf{X}+\lambda \mathbf{I}\right)^{-1} \mathbf{X}^{\top} \mathbf{X} \mathbf{D}_I^{-\gamma} = \hat{\mathbf{B}}_{LAE} \mathbf{D}_I^{-\gamma}, \label{eq:col_item_norm}
\end{align}
where $\gamma$ adjusts the degree of item normalization.

Column-wise item normalization~\cite{abs-1904-13033} solely mitigates the popularity of target items by employing $\mathbf{D}_I^{-\alpha}$ in the first term of $\mathbf{X}\mathbf{D}_I^{-\alpha}-\mathbf{X}\mathbf{D}_I^{-\alpha}\mathbf{B}$ in Eq.~\eqref{eq:dan_objective}.
We also report the performance of column-wise item normalization in Appendix~\ref{appen:norm_comparison}. The results show a performance improvement for tail items, but it does not reach DAN's performance.

\section{Additional Experiments}

\begin{figure}
\begin{tabular}{cc}
\includegraphics[height=2.75cm]{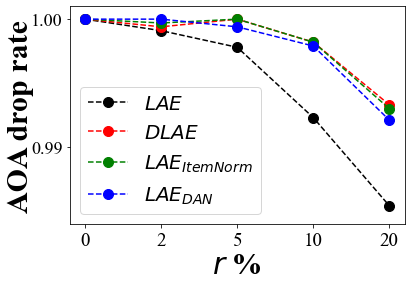} &
\includegraphics[height=2.75cm]{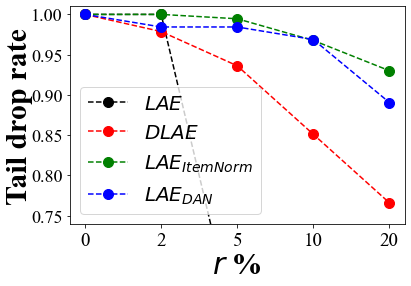} \\
\multicolumn{1}{c}{(a) ML-20M (AOA)} &\multicolumn{1}{c}{(b) ML-20M (Tail)} \vspace{3mm} \\
\includegraphics[height=2.75cm]{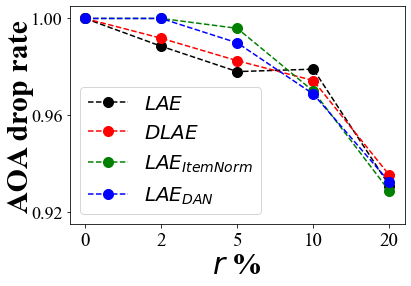} &
\includegraphics[height=2.75cm]{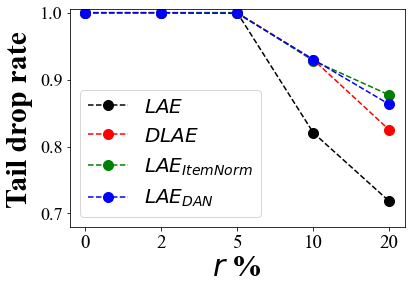} \\
\multicolumn{1}{c}{(c) Yelp2018 (AOA)} &\multicolumn{1}{c}{(d) Yelp2018 (Tail)}
\vspace{-1mm}
\end{tabular}
\caption{Relative performance drop at NDCG@20 on ML-20M and Yelp2018. The y-axis is the decreased ratio of the performance with the noisy training dataset to the performance with the original training dataset.}\label{fig:robustness_eval}
\end{figure}

\subsection{Robustness to Synthetic Noises} \label{appen:robustness}
Following the strategy~\cite{ChenLPWYLZWY22, MaZYCW020}, we evaluate the robustness of item-adaptive normalization of DAN. We randomly replace $r$\% of the observed interactions with unobserved interactions, and we increase the noise ratios $r$ with 0, 2, 5, 10, and 20\%. Figure~\ref{fig:robustness_eval} depicts the performance of the four models (\ie, LAE, DLAE, LAE$_\text{ItemNorm}$, and LAE$_\text{DAN}$) with different noise ratios. (i) The performance drop of LAE models on ML-20M is less than that on Yelp2018. (ii) We observe that DLAE, LAE$_\text{ItemNorm}$, and LAE$_\text{DAN}$ have the lowest relative performance drop in AOA performance. This is due to the denoising ability of item-adaptive normalization, as proven in Appendix~\ref{appen:proof_item_adap_norm}. (iii) In Figure~\ref{fig:robustness_eval}-(b,d), LAE has the largest relative degradation, while LAE$_\text{ItemNorm}$ and LAE$_\text{DAN}$ have lower relative performance drop. This also shows that the denoising effect of DAN is effective for tail items.

\begin{table} \small 
\caption{Performance comparison for existing linear session-based recommender models. `SLIST$_\text{DAN}$' indicates SLIST equipped with our proposed DAN. The best results are marked in \red{bold}, and the second best results are \blue{underlined}.}
\label{tab:session_result}
\vspace{-3mm}
\begin{center}
\renewcommand{\arraystretch}{1} 
\begin{tabular}{P{0.3cm}|P{1.1cm}|P{0.7cm}P{0.7cm}|P{0.7cm}P{0.7cm}|P{0.7cm}P{0.7cm}}
\toprule
\multirow{2}{*}{\rotatebox{90}{}} & \multirow{2}{*}{Model}  & \multicolumn{2}{c|}{AOA} & \multicolumn{2}{c|}{Tail}  & \multicolumn{2}{c}{Unbiased} \\
&    & R@20   & M@20   & R@20   & M@20   & R@20   & M@20   \\
\midrule
\multirow{6}{*}{\rotatebox{90}{Diginetica}}    
& SLIS                  & 0.4988 & 0.1808 & 0.3583 & 0.1206 & 0.0175 & 0.0052 \\ 
& SLIT                  & 0.4311 & 0.1475 & 0.2897 & 0.0830 & 0.0109 & 0.0042 \\
& SLIST                 & 0.4910 & 0.1818 & 0.3919 & 0.1366 & 0.0205 & \blue{0.0062} \\ 
\cmidrule{2-8}
& SLIS$_\text{DAN}$     & \blue{0.5067} & \blue{0.1858} & \blue{0.3943} & \red{0.1497} & \red{0.0224} & \red{0.0066} \\ 
& SLIT$_\text{DAN}$     & 0.4430 & 0.1472 & 0.2943 & 0.0833 & 0.0110 & 0.0045 \\
& SLIST$_\text{DAN}$    & \red{0.5087} & \red{0.1861} & \red{0.3948} & \blue{0.1496} & \blue{0.0223} & \red{0.0066} \\
\midrule
\multirow{6}{*}{\rotatebox{90}{RetailRocket}}        
& SLIS                  & 0.5867 & 0.3625 & 0.5153 & 0.3261 & 0.0488 & 0.0091 \\ 
& SLIT                  & 0.5340 & 0.3106 & 0.4249 & 0.2319 & 0.0301 & 0.0081 \\
& SLIST                 & 0.5907 & \blue{0.3769} & 0.5177 & 0.3244 & 0.0493 & 0.0103 \\ 
\cmidrule{2-8}
& SLIS$_\text{DAN}$     & \blue{0.5939} & 0.3767 & \blue{0.5448} & \blue{0.3667} & \blue{0.0556} & \red{0.0172} \\ 
& SLIT$_\text{DAN}$     & 0.5413 & 0.3239 & 0.4451 & 0.2693 & 0.0322 & 0.0117 \\
& SLIST$_\text{DAN}$    & \red{0.6020} & \red{0.3799} & \red{0.5507} & \red{0.3719} & \red{0.0558} & \blue{0.0163} \\
\midrule
\multirow{6}{*}{\rotatebox{90}{Yoochoose}}     
& SLIS                  & 0.5907 & 0.2580 & 0.2944 & 0.1319 & 0.0047 & 0.0015 \\ 
& SLIT                  & 0.6136 & 0.2809 & 0.3488 & 0.1360 & 0.0045 & 0.0028 \\
& SLIST                 & \blue{0.6332} & \blue{0.2830} & \blue{0.4058} & 0.1812 & 0.0043 & 0.0024 \\ 
\cmidrule{2-8}
& SLIS$_\text{DAN}$     & 0.6317 & 0.2773 & 0.4057 & \red{0.1975} & \red{0.0082} & \red{0.0056} \\ 
& SLIT$_\text{DAN}$     & 0.6204 & 0.2829 & 0.3728 & 0.1517 & 0.0049 & 0.0043 \\
& SLIST$_\text{DAN}$    & \red{0.6375} & \red{0.2909} & \red{0.4174} & \blue{0.1827} & \blue{0.0080} & \blue{0.0049} \\ 
\bottomrule
\end{tabular}

\end{center}
\end{table}

\subsection{Applying DAN to Linear Models in Session-based Recommendation} \label{appen:session_comparison}
To demonstrate the generalizability of DAN, we perform additional experiments in a session-based recommendation environment.

\vspace{1mm}
\noindent
\textbf{Backbone models}. SLIST~\cite{ChoiKLSL21} is the only session-based recommendation model that utilizes a linear item-item matrix\footnote{TALE~\cite{ParkYCL25tale} is a linear item-item model for sequential recommendation, but we did not adopt it as a backbone model because it requires additional temporal information (\ie, timestamps).}. Thus, we utilize SLIST and its two key components, SLIS and SLIT, as the backbone models. SLIS adjusts the diagonal constraints of EASE$^{\text{R}}$~\cite{Steck19EASE} to recommend repeated items, and SLIT organizes the source/target matrix of EASE$^{\text{R}}$ into items that have interacted with the user's past/future to predict the next item. Further, SLIST is represented as a closed-form solution that optimizes both SLIS and SLIT at once.

\vspace{1mm}
\noindent
\textbf{Implementation details}. To adapt DAN to SLIS, SLIT, and SLIST, we modified their objective functions. Since SLIS has the same objective function as LAE, it is applied DAN in the same way as for LAE, \ie, Eq.~\eqref{eq:dan_objective}. For the objective function of SLIT, the user-item interaction $\mathbf{X}$ is divided into two matrices: (i) The past matrix (\ie, $\mathbf{S}$) as input and (ii) the future matrix (\ie, $\mathbf{T}$) as target. Therefore, the diagonal degree matrix $\mathbf{D}_{I}$ is first obtained from $\mathbf{X}$. 
Then, the separated two matrices are multiplied by $\mathbf{D}_{I}^{-\alpha}$, \ie, $\mathbf{S}\mathbf{D}_{I}^{-\alpha}$ and $\mathbf{T}\mathbf{D}_{I}^{-\alpha}$. Since SLIT divides the user-item interaction matrix $\mathbf{X}$ into the past matrix $\mathbf{S}$ and future matrix $\mathbf{T}$, the length of the user history must be calculated differently from traditional recommendation (Eq.~\eqref{eq:dan_objective}).

\vspace{1mm}
\noindent
\textbf{Datasets and Evaluation protocols}. 
We conducted experiments on three benchmark datasets collected from e-commerce services: Yoochoose\footnote{\url{https://www.kaggle.com/datasets/chadgostopp/recsys-challenge-2015}}, Diginetica\footnote{\url{https://competitions.codalab.org/competitions/11161}}, and RetailRocket\footnote{\url{https://darel13712.github.io/rs_datasets/Datasets/retail_rocket/}}. Following~\cite{ChoiKLSL21, ChoiKLSL22SWalk}, we adopted an \emph{iterative revealing scheme} to consider the situation where users repeatedly purchase the same item. We randomly divided the training, validation, and test sets into 8:1:1 on the session side. For evaluation metrics, we adopted two ranking metrics, \ie, Recall and MRR, and also introduced two other metrics to verify the ability to eliminate popularity bias, \ie, Tail and Unbiased evaluation.

\vspace{1mm}
\noindent
\textbf{Experimental results}. 
Table~\ref{tab:session_result} shows the performance of three linear models (\ie, SLIS, SLIT, and SLIST) with DAN. We observe the following two findings: (i) Applying DAN to linear models for session-based recommendation can also improve performance in Tail and unbiased evaluation. In particular, the average gain of MRR@20 in Tail and unbiased evaluation is 10.33\% and 43.13\%, respectively. It demonstrates that DAN can mitigate popularity bias in session-based recommendation as well as in traditional recommendation. (ii) Incorporating DAN into linear models results in an AOA performance improvement. The average gain of MRR@20 on the Yoochoose, Diginetica, and RetailRocket datasets is 2.56\%, 1.64\%, and 3.00\%. Specifically, for SLIST, the performance gain is 2.36\% and 2.79\% on the Diginetica and Yoochoose datasets, respectively. These results demonstrate that DAN appropriately mitigates popularity bias while maintaining or improving overall performance.

\begin{figure}
\begin{tabular}{cc}
\includegraphics[height=2.6cm]{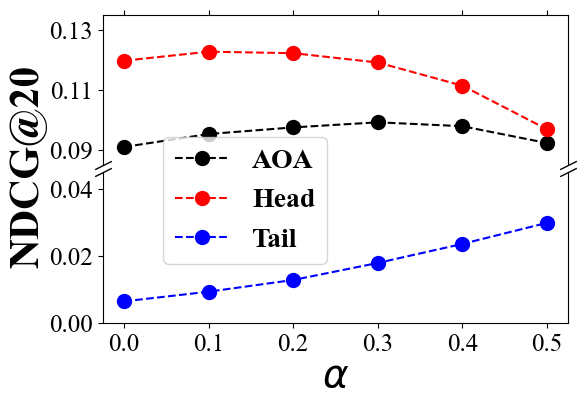} &
\includegraphics[height=2.6cm]{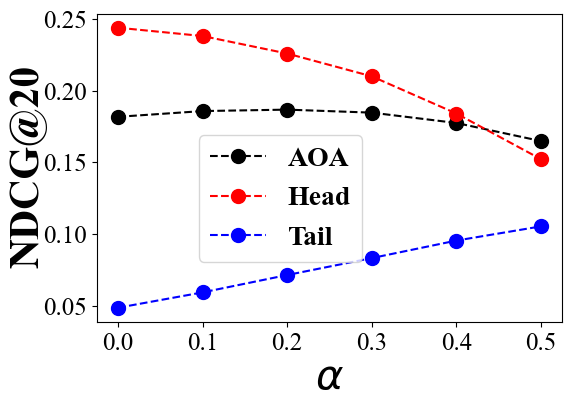} \\
\multicolumn{1}{c}{(a) MSD} &\multicolumn{1}{c}{(b) Gowalla} 
\vspace{2mm}
\\

\includegraphics[height=2.6cm]{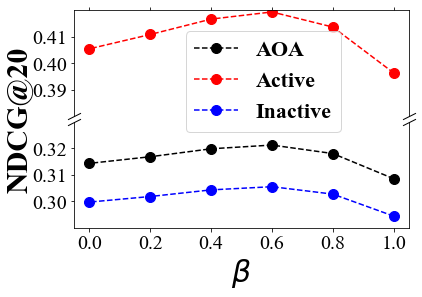} &
\includegraphics[height=2.6cm]{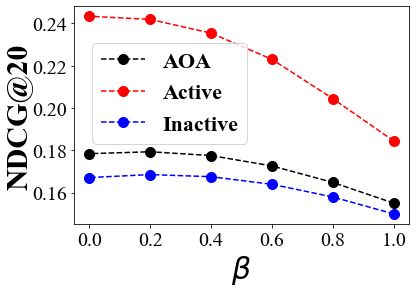} \\
\multicolumn{1}{c}{(c) MSD} &\multicolumn{1}{c}{(d) Gowalla} \\

\end{tabular}
\vspace{-2mm}
\caption{NDCG@20 of LAE$_\text{DAN}$ over $\alpha$ and $\beta$ for user normalization on MSD and Gowalla.}\label{fig:beta_remaining}
\vspace{-3mm}

\end{figure}

\subsection{Hyperparameter Sensitivity} \label{appen:sensitivity}
Figure~\ref{fig:beta_remaining} indicates the performance of LAE$_\text{DAN}$ over adjusting $\alpha$ and $\beta$ for the two remaining datasets, \ie, MSD and Gowalla. Due to their similar $Gini_i$ values, both datasets exhibit analogous tendencies in AOA, Head, and Tail performance with respect to $\alpha$. Meanwhile, Gowalla's higher $\mathcal{H}_w$ value leads to superior AOA performance at lower $\beta$ values than MSD. These findings highlight how the dataset characteristics directly influence the optimal hyperparameters.

\begin{table} \small

\caption{Performance comparison for normalization methods on six datasets with the strong generalization protocol. `Most pop' refers to recommending only the most popular items. `Col norm' refers to the column-wise item normalization~\cite{abs-1904-13033}. The backbone model is LAE. Each metric is NDCG@20. The best results are marked in \red{bold}, and the second best models are \blue{underlined}.}
\vspace{-2mm}
\label{tab:dataset_results}
\begin{center}
\renewcommand{\arraystretch}{1} 
\begin{tabular}{P{1.7cm}|P{1.8cm}|P{0.9cm}P{0.9cm}P{0.9cm}}

\toprule
Dataset   & Method      & AOA         & Head         & Tail                 \\
\midrule
\multirow{8}{*}{ML-20M}
& Most pop       & 0.1355       & 0.1366       & 0.0000             \\
& W/O norm       & 0.3228       & 0.3264       & 0.0001    \\
& Col norm       & 0.3087       & 0.3122       & 0.0023      \\
& RW norm       & {0.3358}       & {0.3386}       & 0.0088       \\ 
& Sym norm       & 0.3321       & 0.3338       & \red{0.0184}      \\
& User norm      & \blue{0.3390} & \blue{0.3428} & 0.0001 \\
& Item norm      & 0.3341 & 0.3379 & \blue{0.0140} \\
& DAN (ours)     & \red{0.3414}       & \red{0.3433}       & {0.0128}         \\
\midrule
\multirow{8}{*}{Netflix}
& Most pop       & 0.1097 & 0.1118 & 0.0000 \\
& W/O norm       & 0.3237 & 0.3316 & 0.0036 \\
& Col norm       & 0.3230 & 0.3281 & 0.0182      \\
& RW norm       & \blue{0.3346} & {0.3358} & 0.0295 \\
& Sym norm       & 0.3307 & 0.3287 & {0.0405} \\
& User norm      & 0.3315 & \red{0.3401} & 0.0010 \\
& Item norm      & 0.3207 & 0.3200 & \red{0.0485} \\
& DAN (ours)     & \red{0.3405} & \blue{0.3388} & \blue{0.0411} \\
\midrule
\multirow{8}{*}{MSD}
& Most pop       & 0.0382 & 0.0447 & 0.0000 \\
& W/O norm       & 0.2740 & 0.2619 & 0.1234 \\
& Col norm       & 0.3030   & 0.2301   & \red{0.2063}      \\
& RW norm       & \blue{0.3204} & \blue{0.2745} & 0.1845 \\
& Sym norm       & 0.2991 & 0.2294 & \blue{0.2061} \\
& User norm      & 0.2549 & \red{0.2804} & 0.0717 \\
& Item norm      & 0.3169 & 0.2630 & {0.1896} \\
& DAN (ours)      & \red{0.3209} & {0.2731} & {0.1873} \\
\midrule
\multirow{8}{*}{Gowalla}
& Most pop     & 0.0219 & 0.0334 & 0.0000           \\
& W/O norm     & {0.1706} & \red{0.2387} & 0.0371        \\
& Col norm       & 0.1822   & 0.2197   & 0.0533      \\
& RW norm      & 0.1693 & \blue{0.2350} & 0.0389         \\ 
& Sym norm     & 0.1637 & 0.1691 & \red{0.0892}           \\
& User norm      & {0.1797} & 0.2317 & 0.0468 \\ 
& Item norm      & \blue{0.1853} & 0.2148 & \blue{0.0798} \\
& DAN (ours)    & \red{0.1911} & 0.2294 & {0.0741}         \\
\midrule
\multirow{8}{*}{Yelp2018}
& Most pop       & 0.0132       & 0.0157       & 0.0000           \\
& W/O norm       & {0.0954}       & \red{0.1285}       & 0.0039        \\
& Col norm       & 0.0963       & \blue{0.1256}       & 0.0084      \\
& RW norm      & 0.0840       & 0.1128       & 0.0044          \\ 
& Sym norm     & 0.0922       & 0.1052       & \blue{0.0217}             \\
& User norm      & \blue{0.0966} & {0.1208} & 0.0101 \\
& Item norm      & 0.0965 & 0.1225 & \red{0.0249} \\
& DAN (ours)    & \red{0.1002}       & 0.1212       & 0.0175          \\
\midrule
\multirow{8}{*}{Amazon-book}
& Most pop       & 0.0085 & 0.0106 & 0.0000 \\
& W/O norm       & {0.1749} & \blue{0.1865} & 0.0635 \\
& Col norm       & 0.1771   & {0.1842}   & 0.0829      \\
& RW norm      & 0.1585 & 0.1668 & 0.0593 \\
& Sym norm     & 0.1627 & 0.1392 & \red{0.0928} \\
& User norm      & 0.1721 & \red{0.1975} & 0.0472 \\
& Item norm      & \blue{0.1782} & 0.1711 & {0.0859} \\
& DAN (ours)    & \red{0.1811} & {0.1700} & \blue{0.0886} \\
\bottomrule
\end{tabular}
\end{center}
\end{table}

\subsection{Comparing Various Normalization} \label{appen:norm_comparison}
Table~\ref{tab:dataset_results} shows the experimental results for the remaining datasets with respect to Table~\ref{tab:dataset_results_two}. It shows the performance of different normalization methods on each dataset. (i) On ML-20M, Netflix, and MSD datasets, RW normalization outperforms W/O normalization, but in the remaining datasets, both RW and Sym normalization degrade AOA performance. However, DAN consistently achieves the highest AOA performance on all datasets, and we can see that the performance of the tail items is also significantly improved. 
(ii) Column-wise item normalization (\ie, Col norm)~\cite{abs-1904-13033} is effective for tail items compared to W/O normalization. It simply mitigates the item popularity bias. In particular, it is more effective on datasets with low $Gini_{i}$ such as Yelp2018.

\begin{figure}
\includegraphics[width=8.3cm]{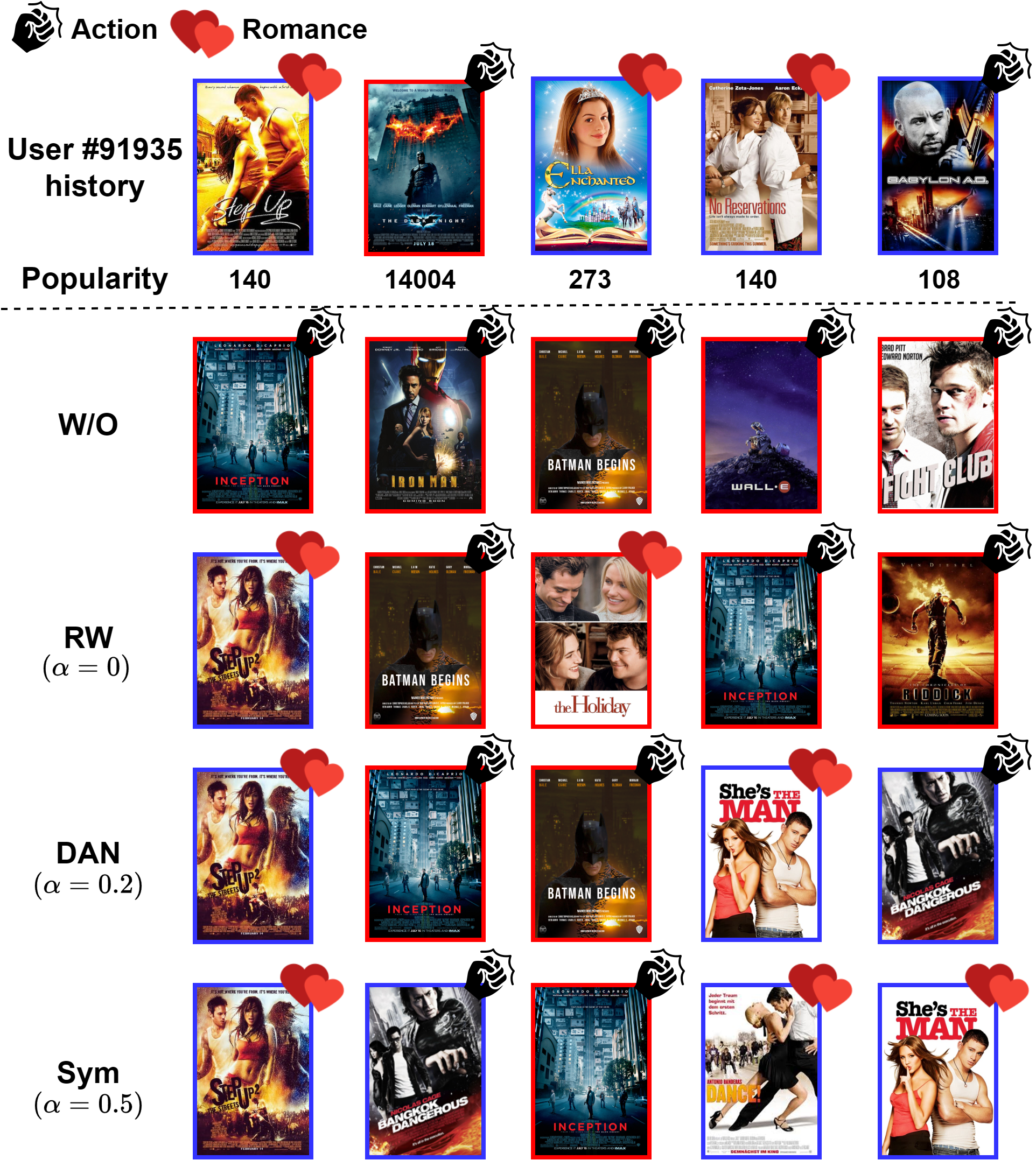}
\vspace{-2mm}
\caption{Case study for various normalization methods on user \#91935 in the ML-20M dataset. Head items are bordered in \textcolor{red}{red}, and tail items are bordered in \textcolor{blue}{blue}.  
}\label{fig:case_study}
\vspace{-2mm}
\end{figure}
\subsection{Case Study} \label{appen:case_study}
Figure~\ref{fig:case_study} depicts the interaction history of the user \#91935 in ML-20M and the top-5 recommendation lists from four normalization methods (\ie, W/O, RW, DAN, and Sym). User \#91935 interacted with movies from both action and romantic genres, including one head item and four tail items. From this case study, we made the following two observations:

\begin{itemize}[leftmargin=5mm]
    \item While W/O extremely recommends five head items, the other three methods recommend tail items appropriately. Even though the user watched three romantic movies out of five, W/O focuses on the most popular action movie \emph{"The Dark Nights"} (its popularity: 14004), recommending five action movies (\eg, \emph{"Iron Man"} and \emph{"Batman Begins"}). The W/O recommendations exhibit high popularity and low diversity, whereas the recommendations from the three normalization methods include tail items from various genres. Notably, all three methods effectively capture user preferences by providing the tail item \emph{"Step Up 2"} (related to \emph{"Step Up 1"} in the user history) as the top-1 item. 

    \vspace{0.5mm}
    \item DAN provides more balanced recommendations by appropriately mitigating popularity bias while maintaining user preferences. RW normalization recommends four head items out of five, indicating that it still does not sufficiently mitigate the popularity bias. In contrast, Sym normalization recommends four tail items out of five, meaning that it excessively alleviates the popularity bias. Unlike the two normalization methods, DAN successfully captures the user preferences and recommends highly relevant items while balancing both head and tail items.

\end{itemize}

\end{document}